\documentclass[11pt]{amsart}
\pdfoutput=1

\usepackage[utf8]{inputenc}

\usepackage[margin=3.25cm]{geometry}

\usepackage{amssymb, amsmath, amsthm, amsfonts}
\usepackage{graphicx}
\usepackage{color}
\usepackage{mathrsfs}
\usepackage{physics}
\usepackage{mathtools}
\usepackage{bm}
\usepackage{tikz}
\usepackage{faktor}
\usepackage{tikz-cd}
\usepackage{bm}
\usepackage[hidelinks]{hyperref}
\usepackage{enumitem}

\DeclareRobustCommand{\SkipTocEntry}[5]{}

\newtheorem{theorem}{Theorem}[section]
\newtheorem{lemma}{Lemma}[section]
\newtheorem{corollary}{Corollary}[section]
\newtheorem{defn}{Definition}[section]
\newtheorem{prop}{Proposition}[section]
\newtheorem{claim}{Claim}
\theoremstyle{remark}
\newtheorem{remark}{Remark}[section]
\theoremstyle{remark}
\newtheorem{example}{Example}[section]

\newenvironment{exenum}
 {\begin{enumerate}[label=\upshape(\arabic*),ref=\theexample.\arabic*]}
 {\end{enumerate}}

\newenvironment{thmnum}
 {\begin{enumerate}[label=\upshape(\arabic*),ref=\thetheorem.\arabic*]}
 {\end{enumerate}}

\usepackage[UKenglish]{isodate}
\usepackage[UKenglish]{babel}

\newcommand{\R}{\mathbb{R}}

\DeclareMathOperator{\E}{\mathbb{E}}
\newcommand{\lag}{\mathcal{L}}
\DeclareMathOperator*{\argmax}{\text{arg\,max}}

\DeclareMathOperator*{\argminemph}{\emph{arg\,min}}

\renewcommand{\grad}{\nabla}

\newcommand{\F}{\mathscr{F}}

\title{Towards a geometry and analysis for Bayesian mechanics}

\author{Dalton A R Sakthivadivel}
\address{\parbox{\linewidth-12pt}{Departments of Mathematics, Physics and Astronomy, and Biomedical Engineering, Stony Brook University, Stony Brook, NY, USA, 11794-3651}\newline}
\address{VERSES Research Lab and Spatial Web Foundation, Los Angeles, CA, USA, 90016\newline}
\email{dalton.sakthivadivel@stonybrook.edu}
\urladdr{https://darsakthi.github.io}
\date{\today}

\let\oldtocsection=\tocsection

\let\oldtocsubsection=\tocsubsection

\renewcommand{\tocsection}[2]{\hspace{0em}\oldtocsection{#1}{#2}}
\renewcommand{\tocsubsection}[2]{\hspace{19.5pt}\oldtocsubsection{#1}{#2}}

\begin{document}

\maketitle

\begin{abstract}
In this paper, a simple case of Bayesian mechanics under the free energy principle is formulated in axiomatic terms. We argue that any dynamical system with constraints on its dynamics necessarily looks as though it is performing inference against these constraints, and that in a non-isolated system, such constraints imply external environmental variables embedding the system. Using aspects of classical dynamical systems theory in statistical mechanics, we show that this inference is equivalent to a gradient ascent on the Shannon entropy functional, recovering an approximate Bayesian inference under a locally ergodic probability measure on the state space. We also use some geometric notions from dynamical systems theory\textemdash namely, that the constraints constitute a gauge degree of freedom\textemdash to elaborate on how the desire to stay self-organised can be read as a gauge force acting on the system. In doing so, a number of results of independent interest are given. Overall, we provide a related, but alternative, formalism to those driven purely by descriptions of random dynamical systems, and take a further step towards a comprehensive statement of the physics of self-organisation in formal mathematical language.     
\end{abstract}

\tableofcontents

\section{Introduction}

The theory of complex systems is, by most accounts, in a Galilean age. Whilst a visionary himself, to whom the shadows of classical mechanics and relativity were known, Galileo's physics was pre-Newtonian. He made observations, and ascribed to them theoretical ideas as a modelling effort\textemdash but, the higher principles organising the phenomena he observed were inaccessible to him, as neither the calculus nor the geometry to fully explain them were known.

Comparably, many examples of complex systems are known to us, and there are effective, if phenomenological, accounts of their dynamics. Biological systems are a particular example of the challenge of studying complex adaptive systems, wherein we are learning increasingly more about underlying mechanisms\textemdash why things \emph{behave} the way they do\textemdash but still know little about organising principles\textemdash why things \emph{exist} the way they do. What largely does not exist yet is a formal approach to the theory of complex systems. This is despite the fact that there are physical foundations for complexity, found primarily in statistical physics, and mathematical foundations for physics, found primarily in geometry and analysis. 

Here, we wish to investigate what this foundational picture might look like. There is a fairly regular assignment of data to any physical theory, consisting of a principle being operationalised by a field,\footnote{By field, we mean an assignment of some quantity to points in some base space\textemdash for instance, values of an electromagnetic force at points in space, generated by electric charges, comprise a classical electromagnetic field. In quantum electrodynamics, this field consists of photon states at every point in space.} a field being modelled by a geometry, and some mechanics as the confinement of a field to the world-line of a system. A field theory is thus a thing which takes a principle and produces a mechanical theory (a prescription of some dynamics) when restricted to an individual system, rather than an entire field. So, if we wish to develop a formal theory of complex dynamics, we might begin by finding the principle and the accompanying field theory that produces such dynamics, and seeing what axioms follow. 

Recent work has suggested that the right principle is an inferential one, established as an account of what it must mean to self-organise\textemdash and thus, to remain far from equilibrium, whence most all features of complexity arise. In particular, the loop between formal mathematics and physical dynamics suggested above has already been partially closed by the free energy principle (FEP) for stochastic dynamical systems, appealing to a mathematical, functional formulation of what complex systems do, using the fundamental optimisation principles known to underlie most of physics \cite{maxwell2018, afepfapp, millidge, fep-simpler}. This is an excellent example of such a mathematical theory of complexity; even so, there are questions surrounding the core articulation and accomplishments of the FEP, many of which remain unresolved (see \cite{aguilera2021} for a key paper in this regard). Laying a geometric foundation for parts of the FEP, and providing a collection of formal structures surrounding it, motivates the results described here. 

Inspired by the revolutionary treatment of physical ideas by formal mathematics, this paper will lay out the beginnings of a theory formulating FEP-theoretic ideas in the language of differential geometry, functional calculus, and category-theoretic elements. 
An emphasis is placed on building the manner in which one might give a formal, mathematical delivery of the free energy principle, as a physical idea relevant to dynamical systems exhibiting features of complexity. 

We begin by giving some remarks about alternative formulations of the free energy principle\textemdash in particular, in Section \ref{duals-subsection}, we sketch out how the free energy of a system's beliefs is the dual function to the entropy of a system's states. We give the context for looking at entropy in Section \ref{constraints-as-blankets}; there, we introduce the idea that constraints in maximum entropy subsume the role of system-ness in the free energy principle, witnessing the equivalence between the two principles. Far and away the longest section is Section \ref{max-ent-FEP-section}, which establishes formally that constrained maximum entropy is an equivalent language in which to discuss the free energy principle. By transforming the free energy principle into a statement about the maximisation of self-entropy, we are given an opportunity to ground the FEP in some conventional mathematics and physics. The remainder of that section makes use of this fact to prove older, and newer, statements about the FEP. 

In the following sections, we use the principle of maximum entropy to establish a field-theoretic language for the free energy principle. In Section \ref{max-ent-field-section}, we discuss how maximum entropy can be thought of as the action of a semi-classical field theory, and in particular, one that interacts with a gauge field. The second major section of the paper is Section \ref{gauge-symmetry-section}, where we discuss previous results on a gauge symmetry in maximum entropy, and show in detail how this is like a gauge force shaping probabilistic dynamics. Important results are given in Section \ref{non-eq-subsection}, concerning how this formalism sheds light on the nature of non-equilibrium steady states. 

More specifically, we will elaborate on three key ideas, rephrasing the ideas about inference put forth by the free energy principle in geometric and analytic terms, and suggesting a formal, proof-based path towards Bayesian mechanics: 
\vskip0.5em
\begin{enumerate}
    \setlength\itemsep{0.5em}
    \item approximate Bayesian inference is equivalent to the maximisation of entropy under a particular constraint (Theorems \ref{ABIL-thm} and \ref{max-ent-FEP-thm}),
    \item this constraint serves as a potential for a system whose description is given by a gradient ascent on entropy (Theorems \ref{constraint-is-potential-thm} and \ref{eff-eq-thm}), and
    \item such a constraint shapes the dynamics of an inferential process in the same way as a gauge field interacting with a matter field (Theorem \ref{funnel-thm}).
\end{enumerate}
\vskip0.5em
Alongside the existing frameworks surrounding the free energy principle, these results move us towards a formulation of the physics of complex systems and non-equilibrium randomness, and advance a view of living things as field theories.

\addtocontents{toc}{\SkipTocEntry}
\subsection*{Acknowledgements}

The author is grateful for discussions with the following people: Miguel Aguilera, Mel Andrews, Lancelot Da Costa, James F Glazebrook, Alex B Kiefer, Magnus Koudahl, Michael Levin, Beren Millidge, and Jeff Yoshimi. The author also thanks the members of the {\tt\href{https://darsakthi.github.io/verses-lab/}{VERSES Research Lab}} for valuable conversations. Special thanks are owed to Christopher L Buckley and Karl J Friston, and in particular, the author thanks Maxwell J D Ramstead. 

\section{Preliminaries}

This paper aims to be at least partially self-contained, and so, overviews of both maximum entropy and the free energy principle are given below. An account of gauge field theory and its mathematical objects is largely left to other resources, with some slightly un-pedagogical commentary in Section \ref{gauge-symmetry-section}. Good references include \cite{rubakov, nakahara, witten}. We assume\textemdash without much loss of generality\textemdash that the systems we discuss and their state spaces are low-dimensional and free of singularities, but not necessarily linear. In key analysis-oriented sections, we assume the underlying noise process is a Wiener process, with some implicit regularity of the operators defined. Throughout, we suppose that we can fully or effectively characterise what it means for a system to be a given system, in the sense of what properties it satisfies, what states it can occupy, and what sorts of dynamics it possesses. Whilst only accessible in principle, invoking the existence of a well-defined `system-ness' stipulating the essential properties of a given system is a necessary basis for any formal investigation of this topic. 

\subsection{Bayesian mechanics and the free energy principle}\label{bayes-mech-prelim-sect}

Self-organisation is the notion that complex systems, especially adaptive systems, seem to exist in stable regimes of states enforced by the very repertoire of dynamics that appear complex\textemdash i.e., non-linear or chaotic\textemdash fashioning a certain order out of disorder. In particular, many features of complexity can be understood as emerging from a sort of synergy between the components of a complex system and their dynamical couplings (see \cite{self-org} for a review). First described in 2006 as a framework in which to understand perception and representation in the human brain as Bayesian estimation \cite{friston2006, friston2010}, the free energy principle is an account of how complex systems become, and stay, organised. It recapitulates several older theories pertaining to perception, learning, autopoiesis, morphogenesis, and the physics of life \cite{fep-simpler}. At first only being applied to the human brain, it has since been expanded to cover a much broader domain, including physical systems in general. It is posited as a comprehensive theory of non-equilibria in stochastic dynamics \cite{afepfapp,parr}. Good overviews of the principle are found in \cite{cocycles-paper} and \cite{maxwell2018}, laying out its mathematical, as well as conceptual, claims; this mathematical content is later examined in detail in \cite{afepfapp}, especially its third section, and exemplified in a mechanical construction found in \cite{dacosta}. Additional worked examples, with commentary on the implied physical dynamics, exist in \cite{aguilera2021} and \cite{stoch-chaos}. An extension of the inference encoded in the FEP to more dynamical situations is effectively summarised in \cite{two-densites}. The most recent comprehensive reviews (at the time of writing) of the FEP are contained in \cite{Andrews2020, millidge, fep-simpler}. Our primary references are \cite{afepfapp} and \cite{dacosta}. This work is in some sense a sequel\textemdash or perhaps more properly, a prequel\textemdash to those two papers, specifically. 
More than a response to critics of the FEP, it is an exercise in clarifying the formal structure and mathematics of the FEP.\footnote{At the time of writing, several responses to \cite{aguilera2021}, a key exposition of weaknesses in the formulation present in \cite{afepfapp}, are to appear in volume 41 of \emph{Physics of Life Reviews}. As such we do not address specific points of critique contained in \cite{aguilera2021}; this will occur in those responses.} 
In effect, we do not aim explicitly to defend the FEP, nor even to patch it\textemdash instead, we assume it is correct, and aim to elevate it, by going back to basics.

The FEP is motivated by an attempt to say what it means for a system to be a system, in the sense of being and staying a cohesive whole. The consequence of this, that an organised system ought to be\textemdash and remain\textemdash independent of its environment, is where the FEP begins its work. As such, it is not exclusively a theory of non-equilibria, but a theory of how structures persist. In particular, it is a theory of what it looks like to remain organised in a stable fashion on some time-scale despite the assault of entropy, and a prescription of what the dynamics of such a system must look like. Whilst this is most informative in the case of complex non-equilibrium systems, whose structures actively resist equilibrium drives, any system with some conception of distinct internal and external states satisfies the FEP in principle. This includes essentially anything that is a stable physical structure (i.e., not a quantum vacuum), and excludes only those idealised systems which are isolated from their environment or embedded in no environment. More pertinently, any structured system which has not lost its structure and joined its environment via some decay can be modelled as resistant to entropy, even if this is only valid on some very brief time-scale, and cannot be reasonably called `resistance' in an enactive\footnote{Roughly, actively or with intention, cf. \cite{two-densites, semantics}. We do not discuss explicit connections to enactive cognitive science here.} sense. The dual viewpoint that any system which is embedded in an environment will reflect the statistical structure of that environment via couplings between the two is equally valid, and seems just as tautological. These points are elaborated on in Remark \ref{intension-intention-rem} and Section \ref{duals-subsection}, respectively. 

We assume a system with internal states $\mu$ is situated in an external environment with its own set of states $\eta$, and couples to this external environment via dynamical interactions across some interface between the two. We further assume that this interface consists of `blanket states' $b$, which are nothing but a distinct subset of system states called a Markov blanket, interacting directly with the environment. Finally, we suppose there is an injective function $\sigma : \mu_b \mapsto \eta_b $ relating a given pair of internal and external states across a shared blanket state. We model the noise inherent in these states as some stochastic differential equation, such that fluctuations in these states are possible; thus, in particular, $\sigma$ maps an expected internal state given a blanket state to the expected external state given that blanket state, and can be thought of loosely as injective on pairs $b \mapsto (\hat\mu_b, \hat\eta_b)$.\footnote{Note, however, that $\sigma$ is not constructed as a tuple-valued function in \cite{parr} or \cite{dacosta}, and it would be a type error to say so; indeed, it is sufficient that $b \mapsto \hat\mu_b$ and $b\mapsto\hat\eta_b$ be injections separately, precisely because $\sigma$ factorises. See Lemma \ref{tensor-hom-lem} for more on this point.} Moreover, interactions of the system with external states are observations of the outcomes of random variables satisfying a stochastic dynamical system, with hidden generative processes. 

It is important to the chain of reasoning in the FEP that a system interfaces with causal factors affecting its states, but cannot `know' those factors directly, and so infers the external states supervening on its own internal states.\footnote{Or can be modelled as such. The entire discussion revolves around this distinction, a not always explicit assumption about the normativity of the FEP; see \cite{Andrews2020} for an account of this problem.} These inferences mean that internal states carry beliefs about the environment, by reflecting the optimal structure mapping onto a particular suite of environmental states; this is true of any embedded dynamical system in at least a trivial sense \cite{bayesian-reasoning}. Suppose that the expected external state $\hat\eta$ is a sufficient statistic for the density describing the environment, or, some variational approximation thereof. We say the states internal to the system play the role of encoding probabilities of external states, such that internal states parameterise a probability density $q(\eta; \sigma(\hat\mu))$ of the likely external states causing particular internal states. This density needs to match a true or optimal belief over states, $p(\eta, \mu, b)$, which occurs when $\mu = \hat \mu$ such that $q(\eta; \sigma(\hat\mu)) = q(\eta; \hat \eta) = p(\eta, \mu, b)$ \cite{dacosta}. From the viewpoint of an agent, it acquires a semantic theory about the content and meaning of external states, which it uses to parse sensory streams\textemdash i.e., how the environment acts on it \cite{semantics}. When this density matches the actual probabilities over external environmental states, the system is in harmony with its environment, and as such\textemdash in the enactive case\textemdash can respond to and resist the sort of environmental fluctuations that cause system dissipation.  

The relationship between internal states, a `recognition density' $q(\eta;\sigma(\hat\mu))$, and an optimal belief model of the environment, $p(\eta, \mu, b)$, is enough to cast the FEP as a theory of inference. Interestingly, in even trivial cases of structures at equilibrium, we have all the ingredients for the FEP: internal states caused by external states, boundaries across which these interactions occur, and most importantly, the statistical structure of the environment reflected in the internal states of the system. We give a case of this in Example \ref{inertness-example}. This suggests we should be able to extract some insights about truly complex systems from much simpler ones in a well-defined fashion, which is a central motivation for our approach\textemdash ultimately, we seek to extrapolate insights about the kinds of systems the FEP aims to describe, from a discussion of simpler systems and the more rigorously understood mathematics of those systems.

We can now offer a short lemma about the `mode-matching' behaviour in the FEP, which was defined above via the synchronisation map $\sigma$.

\begin{lemma}\label{q-mu-and-b-lem}
A system is optimal when it occupies the expected internal state for a given blanket state.
\end{lemma}
\begin{proof}
Suppose the system embodies beliefs about its environment, and carries a recognition density such that a belief about the probability of an external state, $q(\eta)$, is conditioned on a particular internal state $q (\eta \mid \mu)$. We claim a system is able to maintain a non-equilibrium steady state if and only if $q(\eta \mid \mu = \hat \mu_b)$ such that $q(\eta; \sigma(\mu)) = q(\eta; \hat\eta_b)$, which equals $p(\eta, \mu, b)$ by assumption.
\end{proof}

We will sometimes denote $q(\eta \mid \mu)$ as $q_\mu(\eta)$ to emphasise that we are conditioning on an \emph{event}, rather than a random variable, in the interest of using it as a parameter for the sufficient statistics of $q$. This notational choice also appears in the FEP literature, for presumably similar reasons. 

Importantly, the proof in Lemma \ref{q-mu-and-b-lem} is purely heuristic: our assumption in the final line cannot be met, since $p(\eta, \mu, b)$ is formally quite different from $q(\eta ; \hat\eta)$. Nonetheless, we begin what is now identified as the FEP by defining the (also heuristic) free energy:

\begin{defn}\label{free-energy-def}
Free energy is the quantity
\begin{equation}\label{free-energy}
\E_q\big[\ln\{q(\eta \mid \mu)\} - \ln\{p(\eta, \mu, b \mid m)\}\big],
\end{equation}
which is in fact the positive relative entropy between two probability densities over external states, where $p(\eta, \mu, b \mid m)$ is the true joint density over internal states, observations, and their causes, and the density $q(\eta \mid \mu)$ is the belief carried by the system, describing a coupling between external and internal states. The form of \eqref{free-energy} is likened to a KL divergence between the two densities, which is minimised when $q(\eta \mid \mu) = p(\eta, \mu, b \mid m)$ almost surely. 
\end{defn}

In this case, we condition the density $p(\eta, \mu, s)$ on the existence of a random dynamical system $m$, entailing a model of how blanket data is related to external states. This could be thought of as a representation of the blanket, or the model induced by the blanket \cite{cocycles-paper}. Note that the expectation is with respect to the density $q(\eta\mid\mu)$. Due to linearity in the expectation operator, \eqref{free-energy} takes the form of an internal energy less an entropy (cf. cross entropy), and hence is an analogue of thermodynamical free energy. Indeed, it is the energy `available' to the system to enact changes in itself or the environment, and much like a thermodynamical system, stable configurations are tautologically free energy minima, where there is no work to be done. That is, when the free energy of the system approaches zero from above, the system goes from being changed by the environment to being stable. Thus, free energy can be seen as minimised with respect to internal states. Furthermore, the system can decrease free energy by changing its environment, to change the surprisal of blanket states. This comprises a physically-motivated connection of the FEP to organisation and self-organisation. The `internal energy' in $F$ is the average of the intrinsic surprisal of systemic states, so that the surprisal of any such set of states admits an interpretation as a sort of configuration energy. This quantity must be matched by the belief about that state.

We emphasise that the expression in Definition \ref{free-energy-def} is not truly a thermodynamical free energy, but is generically a control-theoretic or information-theoretic quantity \cite{dacosta3, Andrews2020}. We also emphasise (once more) that it is not formally meaningful, but is useful as a motivation for the entire FEP: like other variational inference methods, \eqref{free-energy} is supposedly controlled by how optimal the system's dynamics are. In particular, the internal state $\mu$\textemdash which, we remind ourselves, is in fact an internal state given a blanket state\textemdash is a control parameter for \eqref{free-energy}. Suspending our disbelief temporarily, \eqref{free-energy} is identically zero if $\mu$ is the expected internal state given that blanket state, $\mu = \hat\mu_b$. The blanket dependence of this relationship is a subtle but important feature of the FEP, sometimes obscured by the shorthand notation used here and in other papers. It allows us to define the variational free energy, an approximation to \eqref{free-energy} which \emph{does} make sense:

\begin{defn}\label{var-free-energy-def}
The variational free energy arises from a factorisation of \eqref{free-energy}, relying on an expansion of the joint density $p(\eta, \mu, b \mid m)$ into $p(\eta \mid \mu, b, m)p(\mu, b \mid m)$. Using properties of the logarithm and noting that $-\ln\{p(\mu, b \mid m)\}$ is independent of $\eta$, this results in 
\[
\E_q\big[\ln\{q(\eta \mid \mu)\} - \ln\{p(\eta \mid \mu, b, m)\}\big] - \ln\{p(\mu, b \mid m)\},
\]
such that the KL divergence defined in the first term is a variational approximation to \eqref{free-energy}. The full expression is equivalent to \eqref{free-energy}, and we refer to the divergence herein as the variational free energy. 
\end{defn}

It is important that we take the step to formulate the free energy in terms of a proper KL divergence\textemdash also called a relative entropy\textemdash for various results here. We lose no generality in doing so. Note that in the literature, the full expression above is sometimes called the variational free energy, and thus, so is \eqref{free-energy}. We use the terminology indicated in Definition \ref{var-free-energy-def} for the purposes of results like Lemma \ref{var-free-energy-lem}.

Suppose the blanket consists of sensors of some sort. It is apparent that the variational free energy of a system's observations bounds the surprisal of those observations and resultant internal states, $-\ln\{p(\mu, b \mid m)\}$, from above. Conceptually, the free energy principle can be summarised by the fact that the surprisal of observed states of the environment\textemdash given the model of the environment, $q(\eta \mid \mu)$, encoded in internal states\textemdash is minimised when that model is close to an actual model of the environment. Thus, the principle that `self-organising systems minimise their free energy' is the principle according to which systems remain unsurprised by observations\textemdash in other words, that changes in the environment are predictable, carry little to no new information, and are not allowed to threaten system integrity. Indeed, the expected internal state minimises free energy precisely by mapping to the expected external state, as above. In doing so, we conclude that the probability measure on the state space remains localised around preferable states, the average of which is the expected internal state given some relationship between external and internal states, and take this as self-organisation. 

In summary: mathematically, the minimisation of free energy minimises internal entropy, and thus maximises log-evidence, for the system's own states. It is in this sense that the FEP is essentially a self-evidencing \cite{howhy} of the current and continued existence of a `thing,' since a consequence of the maximisation of self-evidence is that the non-equilibrium probability density over system-like states is resistant to dispersion. That is to say, things that continue to exist\textemdash away from equilibrium, in particular\textemdash are resistant to dispersion, and must therefore be maximisers of their self-evidence. More generally than that, the surprisal given above is a measure of blanket (and thus system) integrity, in that unexpected changes to blanket states register as surprising\textemdash as such, minimising the surprisal of blanket states means the blanket separating internal and external states remains in place.

The active qualities of some of this language obscure the tautology at the heart of their statements, which are generalised from equilibrium systems with structure to describe systems with a stronger sense of adaptivity. Many systems do not assume a model of how the environment causes their states, \emph{per se}, but simply embody evidence of their own existence in virtue of existing \cite{semantics}. In fact, this is consistent with the simplification to the equilibrium case. Equilibrium systems with definite structure evidence by reflecting their environment, which occurs due to the free supervenience of environmental statistics on internal states at equilibrium\textemdash many such systems will even have states of activity, when those internal states affect external degrees of freedom. In that case, trivially, internal states will on average couple to the average external state, but one could describe the lack of adaptivity in these systems as a result of an extremely poor (i.e., inexpressive) model of the external causes of blanket states.\footnote{For example, a stone irradiating heat in the sun may only have one internal degree of freedom tracking external states: its temperature. In the form of synaptic weights, a human may have millions.} On the other hand, systems that remain at non-equilibrium for long periods of time need to `know' what equilibrium is to avoid it\textemdash or, know what fixed point their environment will force them towards, in order to self-evidence by avoiding the changes to internal states intended by the environment. Such systems maintain \emph{a} model, but return to an equilibrium structure, if they fail to minimise free energy for a given set of beliefs. This loss of agency could be described as inertness. In living systems, it is a state known as death. 
See Theorem \ref{change-in-J-info-geo-thm} for more on the `passing down' of Markov blankets. 

The notion of iterated inference as a proxy for changing statistics, such as non-adaptive systems that fail to evidence after a change in the environment, is mentioned in several places. Identifying goal-directed actions with changing the environmental states determining the likelihood of internal states, these actions are an exercise in the preservation of a structure identified with existing internal states. Separately, those internal states can change, to conform or be described by a density of minimum free energy. We will go on to say that the latter is performing inference against a set of constraints on possible internal states.

\begin{remark}\label{intension-intention-rem}
States of activity play an interesting role in the context of the passing down of Markov blankets, underlying a distinction that will be used later in our framework. In standard treatments of the FEP, blanket states are typically separated into sensory and active states, coupled to internal and external states in such a way as to preserve conditional independencies. Via active states, the system can change environmental states. As mentioned, many systems (like a stone irradiating heat) will have trivial active states, and so the distinction between adaptive and non-adaptive systems is more like how those active states are\textemdash or are not, respectively\textemdash used to maintain a given structure in the face of a dynamic environment. On the other hand, having well-defined internal states entails a definition of the system's structure and key properties, and is enforced by sampling from a steady state density of minimum variational free energy; this description fits both adaptive and non-adaptive systems, as mentioned before. We speak more about the example of a stone in Example \ref{inertness-example}, which we refer to as an inert system\textemdash i.e., a system with non-adaptive action states.
\end{remark}

Again, despite the active language used, inert physical systems do not actually \emph{calculate} derivatives. As such, many principles of optimisation\textemdash including the least action principle, the principle of maximum entropy, and the FEP\textemdash instead proceed variationally, such that the optimisation being invoked can be attributed to a gradient flow which might realistically be implemented by a simple object. Many systems\textemdash a diffusing particle, for instance\textemdash are described well by some quantity which always decreases along the spatiotemporal evolution of the system. A \emph{post hoc} interpretation of this phenomenon is that the system is calculating and then flowing towards the stationary point of that quantity, but the truth is much simpler: this quantity is a Lyapunov or $H$-function that controls the dynamics of the system by a basic and quite natural minimisation process\textemdash or even more simply, varies in concert with these dynamics in an insightful way.

As a result, the probability distribution encoded by internal states must flow towards that over external states by some sort of loss-driven dynamics. This is usually spoken of in terms of the KL divergence between the two distributions, which\textemdash in the enactive case\textemdash propagates some sort of error signal to the rest of the system, motivating a self-correction.\footnote{Under a notion of system-ness as consisting of the key properties and states that make a system, identifying ‘entering constrained states as drifting away from allostasis’ with ‘propagation of error by dopaminergic signalling’ recovers the implementation of the FEP in the human brain known as predictive processing \cite{friston2010}. Parallels to other theories like interoception \cite{allen} are also obvious.} For inert systems, the moment error begins to propagate is the moment that structure ceases to exist. For adaptive systems, this gradient flow is the mechanical theory induced by the principle that systems minimise \eqref{free-energy}, in that free energy decreases along the dynamics of a complex system. Indeed, the tendency to minimise free energy is formally a Lyapunov function that any Bayesian mechanical system flows along.

\begin{example}\label{inertness-example}
Consider a stone as an example of an inert FEP system, having an organised crystalline structure, but being at equilibrium and not being enactive. The stone has no dynamics\textemdash there are no goal-directed states of activity originating from internal states, such as rolling away from a threat. Likewise, the stone cannot adapt its internal states to reflect changes in the environment\textemdash a stone will never soften itself to withstand blows of the hammer and remain a stone. All the same, so long as it exists, the stone does minimise its free energy from time-point to time-point, and evidence follows from existence. Here, the internal states of the stone parameterise a set of possible external states (the ambient temperature, for instance), where external states cause internal states and are `observed' by the cortex of the stone. Following Remark \ref{intension-intention-rem}, the existence of this structure is equivalent to system-ness, but the inertness of the stone precludes any enactive or adaptive properties.
\end{example}

In the context of the above example, after being crushed, the description of the stone goes from a crystalline structure to a powder of mineral-based particles. Each such particle has its own internal states and environmental boundary, and so the FEP \emph{still} applies. This is what we refer to as iterated inference, which accounts for the changing statistics of a model of a failed system. Forthcoming work by Friston, Da Costa, and Parr\footnote{See ``the free energy principle: particular kinds and strange things,'' in preparation.} re-derives essential results from the FEP in the absence of Markov blanket assumptions, where the distinction we have made between what sorts of partitions\textemdash active, sensory, coordinated activity, purely environmental\textemdash are possible is also crucial. Like these results, in order to speak of the internal states of something like a stone, it is necessary to specify what sort of content those states could realistically have under the FEP.

Bayesian mechanics should be regarded as a consequence of the free energy principle, in the same sense as classical mechanics is a consequence of the least action principle. Principles are prescriptions of how some theory ought to look; mechanical theories are consequences of a principle, which in turn give us the dynamics of a specific system when applied to that system. In this sense, they are synonymous, as a system minimising its free energy will exhibit Bayesian mechanics. This usage of the term is somewhat idiosyncratic, but is consistent with \cite{dacosta} and \cite{mjdrblankets}, for example. 

\subsection{Inference of distributions and of dynamics}\label{inference-section}

By fashioning a model of a process, inferential things can either consider the probability of an observed state, invert it to estimate the true state in the absence of noise, or more dynamically, solve the master equation associated to a stochastic differential equation. 
Maximum entropy is a procedure for inference which posits that the best model for any data, and hence the best model of (the outputs of) a process generating that data, is that which maximises the differential Shannon entropy subject to prior constraints $J(x)$ on the possible probabilities of states \cite{jaynes, jaynes-book, dill}. Here, by a maximum entropy `model' we mean a probability density encoding the statistics of the underlying noisy dynamics as though a sampling process. This type of model is thought of momentarily as being distinct from what is supposed by the free energy principle, but we will see later that they coincide in a precise sense.

\begin{defn}\label{max-ent-def}
Let $\gamma(x, t),$ denoted $\gamma_t$ for brevity, be an $X$-valued random variable satisfying some stochastic differential equation, with labelled instances $\gamma_t = x_t.$ Moreover, let the process in question generating $\gamma_t$ be stationary, such that $\gamma_t$ is distributed like $p(x)$ for every $t,$ and let $J : X \to \R$ be a scalar measurable function. The Shannon entropy is the action functional
\begin{equation}\label{max-ent-action}
S[p; J] = -\int_X \ln\{p(x)\} p(x) \dd{x} - \sum_k \lambda_k \left(\int_X J_k(x) p(x) \dd{x} - C_k \right), 
\end{equation}
where any $C_k = \E[J_k(\gamma_t)]$ such that the final term is zero.
\end{defn}

In the remainder of this paper, we will forego notating sums, assuming $J$ is a linear combination of $J_k$'s when necessary. We may also write \eqref{max-ent-action} as 
\[
S[p; J] = - \int_X \ln\{p(x)\} p(x) + \lambda J(x) p(x) \dd{x} + \lambda C
\]
to emphasise that there exists a Lagrangian 
\begin{equation}\label{max-ent-lag}
\lag = \big(\ln\{p(x)\} + \lambda J(x)\big) p(x)
\end{equation}
in this action functional, of the more general form
\[
\lag = \big( f(\phi) +  V \big)\phi
\]
for $\phi,$ $f(\phi)$, and $V$ an arbitrary field, function, and potential, respectively. This is used to effect in Sections \ref{max-ent-FEP-section} and \ref{gauge-symmetry-section}, in that it provides a direct analogy to the variational calculus of classical mechanics under the least action principle.

Constraints are usually derived from external knowledge of the data, and often correspond to known values of observables quantities. Inferentially, a constraint is both a function on states weighting the relative probability of a state, and, an expression of some knowledge of what this process ought to look like, as a rule that we know the data must be consistent with. Hence, a constraint is nothing but a synthesisation of relevant information from the data, guiding the likelihood of a given solution. This is reflected in the fact that maximum entropy without constraints is a knowledge-less density, the uniform density, whilst constrained maximum entropy produces a density with no information \emph{other than} the constraints. Indeed, a crucial aspect of maximum entropy is 
that a constraint can be read as a prior specification of what states are likely \cite{constraint-geometry}.  

Due to the way in which \eqref{max-ent-action} is defined, the constraints on states $J(x)$ become moments that the model must satisfy. 
For instance, suppose we know the mean of the random variables we have observed, such that the density describing $Y$ must have expectation
\[
\int_X x\, p(x) \dd{x} = C.
\]
Here, the constraint on a state is proportional to the state, $J(x) = x$.
Maximum entropy can then be constrained such that generated $p(x)$ must produce this given expectation and must be normalised. Let $C_\lambda$ be the rescaled constant $\lambda^{-1} C$ arising from the Lagrange multiplier in \eqref{max-ent-action}. The density generated for this constraint is the exponential distribution with mean $C_\lambda$, given by $C_\lambda\exp\{-C_\lambda x\}$. Indeed, the probability of $x$ decreases as $x$ increases in this density, consistent with the constraint function. 
Denote by $J(x) = j$ the value of $J$ on a state $x$. Maximum entropy provides us the fact that $p(x)$ is a model of the data, and more strictly, becomes a generative model conditioned on the constraints on states, $p(x \mid J = j)$. In fact, it is not just a generative model, but the maximum likelihood model of $\gamma$ given the constraints. Note again that we can take constraints as an expression of the preferability of a particular state, rather than data about the probability density itself; hence we can make sense of $p(x \mid J)$, if $J(x)$ is structured as some observed data about $x$.

Throughout, we make an implicit distinction between inference and dynamical inference. Inference, broadly construed, is the process of finding a model of some data; this is either in the sense of the distribution generating this data, or the distribution from which some process generating the random variables being observed samples. Dynamical inference, on the other hand, ought to fashion a model of the data-generating process as a dynamical system, giving probabilities over the evolution of data\textemdash sample paths, in other words. In virtue of focussing on a simple case of the theory, i.e., stationarily system-like systems with only brief comments on actions, we discuss primarily distributional inference, and possibly iterated distributional inference to encode changes in statistics over time.

Correspondingly, we implicitly refer to the following lemma throughout the paper:

\begin{lemma}\label{stationarity-steady-state-lem}
A process samples from a steady state density if and only if its statistics are stationary.
\end{lemma}
\begin{proof}
Suppose there exists a density $p(x, t)$ describing the system. Both statements are equivalent to $\partial_t p(x, t) = 0$.
\end{proof}
A consequence of Lemma \ref{stationarity-steady-state-lem} is that, all else equal, the difference between an equilibrium and non-equilibrium steady state consists of strictly physical desiderata (e.g., the presence of detailed balance and the nature of the underlying energy flows). 
Indeed, detailed balance is a stronger condition on the existence of a probability density than stationarity, and it is easy to construct stationary Markov processes with asymmetric transitions.
On the physical side, maximising constrained entropy is not truly maximising entropy, leaving the possibility of extracting further work from the system by dissipating it. This difference in entropy is measured precisely by the free energy, the difference between a density $\exp\{-V(x)\}$ and a true $p^*(x)$, as suggested earlier by formulating a free energy as a KL divergence (Definitions \ref{free-energy-def} and \ref{var-free-energy-def}). 

\begin{remark}\label{de-Finetti-rem}
By a consequence of de Finetti's theorem, when the statistics of a process are stationary on some relevant time-scale, the two notions of distributional and dynamical inference coincide on that time-scale. In particular, in the stationary case, each random variable in a temporally-indexed sequence of random variables samples from the same distribution. 
\end{remark}

Using Lemma \ref{stationarity-steady-state-lem} and Remark \ref{de-Finetti-rem}, we take stationary processes as a sufficient fundamental model of an FEP system, as any process which is controlled is stationary, and hence has a steady state dictated by the FEP. In fact, these conditions are also necessary, which is discussed in Section \ref{technical-remarks}. The stability of internal variables in the presence of fluxes at non-equilibrium has been proposed as a hallmark of self-organisation in complex systems in previous literature (see \cite{thermo-complexity} for an extended overview in the context of interacting thermodynamical systems), and the principle of maximum entropy can be employed to describe stable systems in non-equilibrium settings (see \cite{endres} for an explicit construction of such a model and related discussion, \cite{england1, england0, england3} for justifications of coarse-graining non-equilibrium steady state dynamics, or Theorem \ref{eff-eq-thm} in this paper). As such, the resultant `effective equilibrium' of controlled dynamics with a non-equilibrium steady state is used to justify focussing on the simple case of maximum entropy; this is taken as opposed to maximum calibre, a more appropriate generalisation to non-stationary, non-equilibrium, path-dependent systems. Such a generalisation is left to future work. We are able to speak about specific flow properties of some non-equilibrium systems in Section \ref{non-eq-subsection}, but focus strongly on grounding current results and attitudes within the FEP in equilibrium-like cases where we can ignore underlying flows.

If the sufficient statistics $\hat\mu$ of $q(\eta \mid \mu)$, which engage in the mode-matching behaviour described in Section \ref{bayes-mech-prelim-sect}, are fixed, then the density prescribed by the FEP is a stationary density with no goal-directed action component. This precludes a description of intentionally evolving systems, as suggested above. Action can now be introduced as non-stationarity\textemdash a temporally-indexed change in these statistics deriving from a change in the matched external state. This is a signature of an iterated maximum entropy process, describing a system whose steady state might suddenly change, rather than a highly controlled process. Note that this non-stationarity appears to include adaptive systems \emph{and} inert systems, both of whose embodied evidence changes when the system changes (undergoes changes in, respectively) its form. This means we have partitioned the class of systems into four subclasses under the FEP, two of which are stationary (experience no \emph{effective} energy flows), and two of which are adaptive (have active states), where no such subclass is empty.

\begin{example}\label{classif-ex}
We provide examples of the sorts of systems contained in each subset here.
\vskip0.5em
\begin{exenum}
    \setlength\itemsep{0.5em}
    \item A stone at a single point in time is a stationary, inert process. \label{stat-inert-ex}
    \item A control circuit is a stationary, adaptive process. \label{stat-adapt-ex}
    \item A stone over multiple points in time is a non-stationary, inert process. \label{non-stat-inert-ex}
    \item A life-like system is a non-stationary, adaptive process.\label{non-stat-adapt-ex}
\end{exenum}
\vskip0.5em
\end{example}

The difference between stationary and non-stationary adaptivity (Examples \ref{stat-adapt-ex} and \ref{non-stat-adapt-ex}) is whether or not the attractor characterising system-like states wanders: is the system changing its environment to maintain a fixed, stable regime of states, or, does it change its statistics in concert with the environment? In contrast, the difference between inert and adaptive stationarity (Examples \ref{stat-inert-ex} and \ref{stat-adapt-ex}) is whether the system is resisting equilibrium drives or not. We have also introduced systems which are non-stationary and inert. We could contrast Examples \ref{non-stat-inert-ex} and \ref{non-stat-adapt-ex} by asking whether any changes in the system originate from the system, or if they are all environmental. These are the sorts of systems that the FEP only applies to vacuously\textemdash dissipating systems that change because they can't help themselves\textemdash but which nonetheless can be discussed in the language of the FEP, because on some time-scale they appear like Example \ref{stat-inert-ex}. The difference between the two could be characterised as a sort of meta-stability, or stability in the face of environmental perturbations predicated on appropriate active responses.

Concomitantly, we take the position that the FEP\textemdash as written, or perhaps re-written, here\textemdash speaks of system-ness in the broadest sense possible, covering things which are both at and out of equilibrium from time-point to time-point. We focus on systems like Example \ref{stat-inert-ex}, generalising to systems like Example \ref{stat-adapt-ex} where possible. As stated, the point of this is to suggest what mathematically well-understood features can be expected of more general systems still, like those of Example \ref{non-stat-adapt-ex}. Since systems like Example \ref{stat-adapt-ex}\textemdash other examples include systems like Turing patterns, reaction-diffusion systems, and the morphogenesis of differentiating cells\textemdash can be regarded as performing inference \cite{kuchling, fields}, termed `cognition in unconventional substrates' in \cite{levin}, this is an approach that offers ready insights to life-like systems. Moreover, as we will discuss in Section \ref{constraints-as-blankets}, even living systems like humans have certain static bounds on what constitutes such a system. This should allow us to fashion models of life-like things using the formalism defined here, albeit coarse ones which may not be very informative.

\subsection{Dualities in categories and in inference}\label{duals-subsection}

An \emph{adjunction}, as found in category theory, is the existence of a pair of maps which are adjoint; that is, one is dual to the other. Dualisation maintains key intrinsic properties of objects, but reverses the directions of the relationships between objects. Dual maps are thus precise opposites: each map is one of two sides of a shared coin. The relationships encoded by adjunctions are of critical importance in mathematics, especially via the application of universal properties, and most every non-trivial mathematical structure arises from a duality of this sort \cite{lawvere, maclane, awodey}.\footnote{Indeed, the stated slogan of \cite{maclane} is ``adjoint functors arise everywhere.'' Mac Lane also makes the claim that ``all concepts are Kan extensions,'' which come in left and right dual pairs.} 

In this spirit, within the free energy principle there is a symmetry between a system and the environment embedding it, referred to as `synchrony' across a Markov blanket \cite{fep-simpler}. This is evident in the way environments tend to model agents just as well as agents model their environments, a kind of niche construction \cite{niches}. Moreover, the FEP naturally induces a so-called `dual information geometry' by defining the simultaneous evolution of internal states and external states \cite{parr}. This allows us to speak intelligently about the agent in the context of the coupling between agent and environment, or more generally, about agent and environment as systems related by a coupling (see \cite[section 4.2]{semantics}). A non-trivial formulation of the bidirectionality of such interactions (often focussing on the flow of relevant quantities) has been essential to other rigorous discussions of open systems \cite{rosen1, yoshimi, yoshimi3, siekmann, thermo-complexity,  bayesian-reasoning}, especially those rooted in category theory \cite{rosen2, fong, baez-nets, petri, courser, matteo, toby, master, fibre-optics}, suggesting it should play a \emph{key} role in understanding the FEP. In fact, formulating systems with bidirectional information flow necessitates an explicit adjunction between two interacting agents\textemdash a covariant and contravariant pair of morphisms \cite{hedges}\textemdash further suggesting that we should explore this symmetry from within the FEP as a dualisation of agent into environment.

In summary, the fact that a Markov blanket encodes a kind of adjunction\textemdash between which states track which\textemdash suggests that a given (outward- or inward-facing) dynamic across a Markov blanket is ultimately the dual of some other, adjoint dynamic, described by facing in the opposite direction. The content of this paper is in large part preoccupied with what this dual perspective has to say about the problem of self-organisation, and what value it may have as a technical and conceptual tool. 


The FEP works by focussing not on the system, \emph{per se}, but on the coupling between the environment and the system, depicting what environmental forces look like from the perspective of the system.\footnote{It would be appropriate to note here that a Markov blanket is not strictly a \emph{separation} of the system from its environment, despite what appearances might have us think. Rather, it is a coupling between two distinct subsets of an entire locale of states, with one being a cohesive whole that we call the system. This distinction is critical to the philosophical effectiveness of the FEP, in that it does not falsely presuppose that open embedded systems make themselves into closed systems; this is not even possible in principle, and in fact, the FEP \emph{leverages} this non-closure to define a mechanical theory predicated on the coupling. See also Theorem \ref{joint-entropy-thm}.} The FEP has a natural control-theoretic interpretation as a statement that the dynamics of a system can be described as responses to perturbations from the system's surrounding environment (this is present throughout the literature, but is arguably exemplified best in \cite{control-as-inf}, a detailed account of the similarities between control and temporally-extended FEP-type inference), with a system's dynamical repertoire being more akin to a set of tools for coping with a variable environment, rather than a defining functional or phenotypical feature of the system. It is a consequence of this framework, favouring the viewpoint of the environmental states being predicted or reacted to, that the dynamics of a system can be understood as minimising free energy. In other words, the FEP is the principle prescribing what the dynamics of a complex system ought to look like, in that the maintenance of a steady state far from equilibrium requires an active flow against some gradient. Under this framework, the system embodies or even carries a generative model of environmental states, simply by representing what it is to exist in a given environment. In this sense, embedded dynamical systems in general are like agents of inference: we can, in principle, `read off' the beliefs about the environment encoded by the states of a system, and thus, the semantic content stored in the internal states of the system (see \cite{bayesian-reasoning} for an example). This is another facet of the same symmetry: as outside observers situated in the environment surrounding the agent, \emph{we} can fashion a model of how the agent behaves, which is dual to the model carried by the agent about how the environment behaves. Dualisation introduces the agent as an entity which exists from the point of view of the universe, by asking what it means for an agent to be a cohesive region of states distinguished from its surroundings. This idea is introduced formally in Section \ref{constraints-as-blankets} as a set of constraints on what the system can be, such that constraints enforcing a systemic ideal are adjoint to a system having ideal explanations of its environment. Being equivalent to the principle of maximum entropy as it was described in Section \ref{inference-section}, this move allows us to speak about constrained self-entropy rather than the free energy of beliefs. In this sense, dualisation is little more than a technical tool that relates the FEP to the entropy functional, but it offers an attractive alternative viewpoint on the FEP.

Spelling out the formulation of this adjoint pair in greater detail, we have the following: take a set of possible parameter values for $q(\eta;\mu)$ equivalently as defining a conditional distribution $q(\eta \mid \mu)$. Formally, the surprisal is an evidence bound for the model $q(\eta \mid \mu)$. That expected internal states minimise \eqref{free-energy} prescribes that non-equilibrium systems must create and climb positive probability gradients, 
thereby staying self-organised by assuming a model concordant with what it is to exist in a particular external environment. This is the density $p(\eta \mid \mu, b)$ that FEP-type systems embody by way of internal states, where those states model some belief about external states via a map across the Markov blanket; it corresponds to estimating likely external states given what the system looks like, which is how the system remains system-like in the face of a changing environment. The minimisation of variational free energy is precisely the leveraging of this map to make embodied beliefs model true descriptions of the environment. 

This says little of self-evidencing, though, and could (albeit na\"ively) be construed as the opposite: we understand systems more directly as things which model their environments, rather than things which collect evidence about themselves; it is only the fact that the latter is a consequence of the former in the variational case of the FEP that evinces self-evidencing. Moreover, there is immediately a suggestion that the Bayesian inversion of $p(\eta \mid \mu, b)$ is the more canonical viewpoint. In a sense, we are secretly asking about what a system probably looks like, or is allowed to look like under a mutual understanding of the defining qualities of that system, conditioned on what the environment looks like\textemdash and this is $p(\mu\mid\eta_b)$, where we have absorbed blanket states into internal states. From a certain point of view, \emph{this} is what really defines self-organisation, in that we are directly estimating the probability density of states given the aforementioned external perturbations to those states. Initially, it seems like this flips the entire story on its head. Now, we \emph{are} interested in the dynamical repertoire of the system for what it says about the system, whilst we are \emph{no longer} interested in the generative model encoded by the system. We have changed our perspective to focus directly on what the system looks like from the perspective of environmental forces. And yet, in both contexts, the tendency to minimise free energy is a means to understand system-ness, whether by our own model or the system's model. The fact that the minimisation of free energy is recognisable in both `narratives' is, in fact, a key feature of the FEP. Our intuition is clearly wrong: this new interpretation is not so radically different from the original, even if its mathematical formulation has reason to change.

To determine what this dual formulation\textemdash focussing on what internal states look like from the perspective of the bath they are situated in\textemdash ought to look like, we can exploit the symmetry of the Markov blanket. The adjunction between external states supervening on internal states, and internal states responding to external states, is a very literal\textemdash in the sense of Bayesian inversion over a Markov blanket\textemdash duality between system and environment. This paper takes the route of formally establishing that maximising \emph{constrained} entropy over internal states tends to minimise free energy, such that localising the probability density around optimal states implies the minimisation of the free energy of internal states. We are able to prove this largely by relying on that duality, the foundation for which is established in the following section.

\section{Constraints as specifications of system-like states}\label{constraints-as-blankets}

A section devoted to the special role constraints play in self-evidencing is contained here, preempting the technical step that makes many of these results possible: passing from free energy to self-entropy, which requires us to formulate the problem in terms of constraints. Constraints are most easily understood in terms of preferences for certain internal states, perhaps reflective of the naturally adjoint structure of the problem.

We are principally interested in what the evidence of self-ness is, or, what is the actual definition of a system that a system can be understood as meeting by existing. 
For a system to be a system means that it obeys its own `system-ness'\textemdash whilst initially nonsense, what this means is that there are a small number of characteristically system-like states, which are occupied with high probability, such that the system is and remains whatever sort of system it is. This is what we refer to using terms like organisation and self-organisation\textemdash that the system exists means that it remains in a regime of states which give it the properties it has. This is what we term an `ontological potential,' defining a set of states with key properties. By falling into an ontological potential, a system is organised into occupying system-like states. 




In turn, this is quite closely connected to the fact that observations of unexpected environmental states can cause transitions to unpreferable internal states, compromising the system’s system-ness. In adaptive systems, self-evidencing entails enforcing this distinction\textemdash and thus persisting on a long time-scale\textemdash by modelling the environment embedding the system, and then responding to those changes. Even in the adaptive case this contains the rather more trivial, special case that systems specialise to exist in the environment they exist in. This is certainly a tautology, and itself includes the simple structure-at-equilibrium case, where a system takes on the statistics of its environment via uninhibited energy flow. As in Example \ref{inertness-example}, a stone does not \emph{self}-evidence, but it does evidence, in virtue of existing with a `model' that captures the free supervenience of external states on its own states. It would begin to evidence an entirely new set of states were it to get smashed to pieces (again, we point the reader to Theorem \ref{change-in-J-info-geo-thm} for a formalisation of this).

Essential to the definition of system-ness is obviously a set of constraints being placed on what it means to exist as a thing. For humans, this consists of allostatic set-points comprising healthy system dynamics\textemdash the amount of energy, fluids, and waste product in the body, for instance. These shape the dynamics of the thing and enforce system-ness when they are obeyed\textemdash eating when hungry, staunching bleeding when injured, excreting waste when it builds up. 
Constraints can thus be read as preferences, in that states with low constraints are `good' states for the system. Given some notion of system-hood\textemdash i.e., the key variables being constrained\textemdash we can think of constraints as inducing an attractor for the system, in that they define a region of states that a system can sample from or occupy and still be like that system (under some fundamental notion of what sort of system it is). The constraints on the system are the ontological potential defining how we might model a system. Crucially, it is also not necessary to have the entire essence of a thing written down; we need only define a collection of key internal state variables which are independent of key external state variables. These include what are termed `existential variables' in \cite{Andrews2020}, which manifest self-evidencing by a sort of contradiction\textemdash these are the variables for which the system surely dissipates if not kept in allostatic bounds, or in other words, any variable whose obedience of some bound is necessary in order for the system to be defined. The constraints on these quantities constitute the potential, and the maintenance of their values in a regime which is amenable to self-organisation is allostasis.

As seen in maximum entropy, constraints on what states are accessible to a system enforce a probability density taking into account the relative probability of such states. Hence, any state sampled by the system can be understood as occuring with a probability dictated by the constraints on that system\textemdash definitionally, systems do not sample non-system like states, but there may be some flexibility within the allostatic bounds mentioned before. Maximising entropy thus gives us a route to claim that internal states model external states, contingent on pre-specifying the desired internal states constraining the system.

The viewpoint of constraints demands something of a reinterpretation of the FEP, which discards the more computationally useful Markov blanket, invokes some platonic representation of what it is to be a given system, and then claims that we can understand a systems’s system-ness in light of maximising entropy against system-ness. This becomes evidencing, in that we begin with what the system is and ask how it is that way\textemdash or self-evidencing in the adaptive case, where we begin with what the system wants to be and ask how it gets there. In other words, ours is consistent with the recent, more deflationary view of the FEP, in that we presuppose what the system ought to be and then describe how it attains system-ness \cite{semantics, Andrews2020}. 

\begin{remark}\label{variance-constraints-remark}
It is interesting that many such constraints will be variance constraints, or, a $J$ function which is quadratic in $x$, in the sense that states which are too high \emph{or} too low ought to be penalised equally under the definition of a controlled system. Occupying these states, which lie outside of some sort of set-point, for long periods of time would certainly violate negative feedback principles in systems biology \cite{wiener}. In this case, the resulting maximum entropy probability distribution would be of a Gaussian shape, and the idea of `peaking' around system-like states is taken literally. This lends a concrete, control systems interpretation to the observation that increasing the variance of $p(x)$ and sampling more surprising states is the dispersion of the system's system-ness. The physical role of free energy minimisation as a definition of system identity is also discussed in detail in \cite{kiefer2020psychophysical}.
\end{remark}

Much has been made of the Markov blanket utilised in the FEP literature (see \cite{MB-trick} for a detailed critique). The viewpoint emphasising constraints appears to avoid some of these problems, at the expense of being less useful for general calculations. It will be apparent in Theorem \ref{max-ent-FEP-thm} that the Markov blanket plays at most an auxiliary role, and the constraints imply a boundary between system and environmental states without referencing the statistical structure of either. This is, in fact, an advantage to our approach: consider that anything which persists as a thing falls vacuously under the FEP, and that Markov blankets ought to be just as fundamental as a result. Note that in the following, we mean well-defined in the technical sense of a property which is independent of some choice.

\begin{prop}\label{MB-prop}
There is no well-defined, maximal proper subclass of stationary embedded systems with the property that each such system is equipped with a Markov blanket.
\end{prop}
\begin{proof}
We suppose that every embedded system either has a Markov blanket or does not, and assume there exists a generic thing with fixed internal states which admits a representation of those states that is \emph{not} conditionally independent of external states. In other words, we assume a distinguished subclass of blanketed systems exists, which is neither empty nor equal to the entire set of stationary embedded systems. Now, for that generic thing, choose a partition of the statistically fixed internal states into blanket states. By including sufficient information in the form of latent blanket states to predict internal states independently of external states, a Markov blanket can be constructed for that system. The claim follows by contradiction.
\end{proof} 

Hence, for any collection of internal states\textemdash even a stone\textemdash we may construct a Markov blanket on those states. This proposition is a special case of the more general fact that any system can be made Markovian by conditioning states on adequately many hidden variables. This may be one explanatory variable or many; either way, conditional independence is somewhat arbitrary, and some sort of blanket can be constructed for arbitrary systems.\footnote{Despite appearing to be an esoteric statement, it is actually fairly straightforward: all non-Markovian dynamics admit a representation as a hidden Markov model, and once we include hidden variables as state information, the system is made conditionally independent of some other quantity. A good example is that internal states are independent of external states when conditioned on the variable of control. Since we assume control equates to the presence of a Markov blanket, stationarity implies a Markov blanket.} Now, note how we began this line of reasoning: declaring there exist distinct internal states within some bath characterising the system of interest implies a coupling that automatically leads to some manner of conditional independence. As such, constraining the system to remain in a certain regime of internal states that allows those states to cohere and persist is equivalent to asking that a Markov blanket exist between internal and external states.\footnote{Note how this proposition does not tell us how to find the Markov blanket for any system, nor if it is meaningful from the point of view of physical calculations\textemdash Proposition \ref{MB-prop} is merely an existence proof, as opposed to an algorithm. A choice of what counts as a useful condition for the Markov blanket is the element of well-definiteness given in the statement, in that more restrictive ideas of a blanket can be introduced that carry stricter conceptual content.} 

It must be stated again that this restores the validity of the FEP as a foundational approach to every `thing,' since it is in fact the case that distinct internal states of any kind ought to be sufficient for a boundary of some sort. As a further remark, note that \eqref{max-ent-action} has no preferred length- or time-scale\textemdash since the action is scale-free, the theory applies to the dynamics of most any system. The Markov blanket construction itself is scale-free, as has been described in \cite{particles-and-parcels} using a renormalisation-type method; dually, any system can certainly be constrained to be system-like.

On the other hand, it preserves the fact that the FEP is not so general as to be useless. Whilst these trivial Markov blankets exist, they are nothing interesting. Inert systems, like the crystalline structure of a stone, are not controlled (see Example \ref{non-stat-inert-ex}), and so do not truly have a Markov blanket. Proposition \ref{MB-prop} is the sort of sleight of hand that allows us to talk about a simplified version of the FEP; in reality, the FEP has a far less trivial heart that applies only to very particular systems with complexity on the order of Examples \ref{stat-adapt-ex} or \ref{non-stat-adapt-ex}. The Markov blanket is thus more special than Proposition \ref{MB-prop} would lead one to believe. 

An interesting case study for this construction, and in particular, the fact that generalisations of constraints generalise blankets, is that of a flame. Consider one of the key issues raised in \cite{MB-trick}: that of a `wandering blanket.' This is the idea that an evolving definition of a system, especially a system whose states fluctuate too quickly to maintain the renormalisation-type relationships yielding the statistical independencies defining the Markov blanket, fail to accomodate a Markov blanket. The example often used is a flame, which converts itself from a fluctuating plasma-like substance to carbon gas too quickly to have a blanket. However, in the process of discussing why the Markov blanket fails, we have just sketched out what the flame is and what the flame is not. Evolving systems more generally have a formal home in the notion of dynamical constraints, given that internal states can be defined or approximated by a dynamical system at some scale of observation with some salient state variables, no matter how quickly that changes. We might, for instance, model a flame's essential flame-ness as the maintenance of a temperature above the ignition point of the substance being burned, a non-zero flux of oxygen around the material, as well as the occurrence of some particular chemical reactions related to combustion, without any one of which there is no flame. These are the existential variables for the flame. Suppose the flame gets snuffed out; we have moved the flame to non-flame-like states, and this can be regarded as a failure to evidence. The fact that the flame is not adaptive is captured by the flame not self-evidencing, and simply \emph{being} evident by existing. That is to say, it does not \emph{enforce} its own existence. The fact that the state of the flame fluctuates within these constraints is captured by the fact that the maximisation of entropy around said constraints provides a certain amount of exploration in the state space. In this case, maximum entropy is a \emph{post hoc} description of the system's continued existence, rather than a mechanism for self-preservation, so to speak. 

A key to the distinction above is that we don’t need to consider the difference between evidencing and self-evidencing: from the perspective of the environment, these look the same. In both cases we are referring to a coherent collection of things which together define a particular system. In the absence of perturbations from the environment, these notions coincide. Much will be made of this mental framework for the internal states of an agent, in that this view of constrained system-ness underlies the entire construction in this paper.

Dynamical constraints are not discussed here, but follow from the framework laid out, and\textemdash conjecturally\textemdash solve issues related to non-ergodicity in itinerant systems with changing blankets. See Theorem \ref{ergodic-thm} for the relationship between constraints and ergodicity, and Theorem \ref{funnel-thm} for the relationship between constrained ergodic densities and attractors, which together suggest that dynamical constraints will yield a rigorous notion of wandering attractors. The adjoint view\textemdash that systems can apply dynamical constraints to their models of the environment\textemdash has been described in the literature before \cite{yoshimi2}, and may be related to the expected free energy construction \cite{gefe, parr, efe}. As such, this is a promising future direction for the modelling of genuine non-equilibria.

\section{Optimisation as self-evidencing}\label{max-ent-FEP-section}

The content of this section first establishes how the principle of maximum entropy subject to a particular constraint is equivalent to the free energy principle, in that it leads to self-evidencing of self-organising systems (or an evidencing of organised systems) and induces an information geometry on a system's possible models. The equivalence between these theories is best understood in light of constraints on the possible states of the \emph{system}, motivating the dualisation described in Sections \ref{duals-subsection} and \ref{constraints-as-blankets}. We go on to leverage this new construction, using maximum entropy to discuss some formal mathematics for self-organisation in Sections \ref{geometry-and-analysis-sect} and \ref{technical-remarks}. In Section \ref{as-if-inference-subsec}, we zoom back out, to discuss what maximising entropy entails in a philosophical, agent-based context.

Recall that a maximum entropy probability density is a representation of its constraints (see Section \ref{inference-section}). We assert the following claim:
\begin{claim}\label{FEP-is-max-ent-claim}
The right interpretation of the FEP is that non-equilibrium dynamical systems are data-generating processes whose data is the maximum likelihood model of the system given the constraints on what that system can be. This likelihood is self-evidence, and maximising likelihood is self-evidencing. As such, FEP-type systems maximise constrained self-entropy.
\end{claim}
This claim leads to various features of complexity, by way of non-equilibrium dynamics, purely as a result of representing and responding to constraints. 
It can be read as a statement about 
nestedness\textemdash anything that exists, exists in some environment, and is a microcosm of that environment in virtue of reflecting just what it is to exist in that environment. As such, we can model a system as a thing in an environment which is constrained not to join that environment. 

Motivating Claim \ref{FEP-is-max-ent-claim} is the observation that, as well as its important roles in probability and physics, the maximisation of entropy underlies much of machine learning. Featuring in Boltzmann machines \cite{boltz-mach} and forms of energy-based learning \cite{ebl, gan}, physical descriptions of inference lend themselves to entropic descriptions in particular. The `metalogical' aim of this paper is also to show that the FEP is true whenever maximum entropy under a particular constraint is true, simultaneously demystifying it and giving it a solid foundation.


Expounding upon Claim \ref{FEP-is-max-ent-claim}, we obtain the construction we need, as follows. Recall that minimising free energy implies the presence of an attractor of states\textemdash a region where the probability of occupying any state contained therein is high. Saying the dispersion of this probability density is low, and hence it is low entropy, is ultimately misleading: in reality, we maximise entropy under a certain constraint against dispersion, leading to a high probability of occupying states in that region. Moreover, minimising free energy does not construct that attractor explicitly, but returns the internal states that parameterise an optimal belief about the environment.

For these reasons we pass to constrained entropy over beliefs and then constrained self-entropy, a transformation happening in two steps: first, we equate the minimisation of free energy to the maximisation of a particular entropy functional, and then we use the adjunction sketched out in Section \ref{duals-subsection} to change our perspective. This will allow us to relate the self-evidencing aspects of the FEP to the implicit attractor on states, and, allows us to invoke results on the entropy functional to understand the FEP. 



We denote by $q_\mu(\eta)$ the conditional density $q(\eta \mid \mu)$. To begin, observe that maximising the first term in \eqref{free-energy} minimises the overall expression. This motivates us to choose the $q_\mu(\eta)$ for which $S[q_\mu(\eta)]$ is greatest, under the constraint that $q_\mu(\eta)$ characterises its surroundings well.

\begin{prop}\label{max-ent-q-prop}
For a fixed density $p(\eta, \mu, b \mid m)$, the minimisation of \eqref{free-energy} occurs if and only if $S[q_\mu(\eta)]$ is maximised under the constraint that the optimal belief about the agent-environment loop is on average unsurprising, $J(\eta) = -\ln\{p(\eta, \mu, b \mid m)\}$ and $\E_q[J(\eta)] = 0$.  
\end{prop}
\begin{proof}
Negating \eqref{free-energy} and rewriting it as
\begin{equation*}
-\int \ln\{q(\eta\mid \mu)\} q(\eta \mid \mu) \dd{\eta} + \int \ln\{p(\eta, \mu, b \mid m)\} q(\eta \mid \mu) \dd{\eta},
\end{equation*}
by the logic of \eqref{max-ent-action}, we have a maximum entropy problem. Since maximising this expression is equivalent to minimising its negation, we have the intended result. 
\end{proof}

It should be noted that we could have constructed this relationship via the Legendre transform, but that this is largely unnecessary. A particular statement of one direction of Proposition \ref{max-ent-q-prop} should be made:
\begin{corollary}\label{q-equals-p-corr}
Maximising $S[q_\mu(\eta)]$ under $J(\eta) = -\ln\{p(\eta, \mu, b \mid m)\}$ implies the optimal probability density $q^*_\mu(\eta)$ is equal to $p(\eta, \mu, b \mid m)$.
\end{corollary}
\begin{proof}
This follows from maximum entropy, which yields $q^*_\mu(\eta) = \exp\{-J(\eta)\}$.
\end{proof}

A consequence of Proposition \ref{max-ent-q-prop} is that minimising particular free energy as written necessarily implies that internal states model external states. Likewise, this unique constraint is sufficient for a maximum entropy density to describe a system modelled by the FEP in the context of particular free energy, in that it implies 
the expected surprisal of external states under $q$ is zero. Both are summarised effectively by Corollary \ref{q-equals-p-corr}, $q^*_\mu(\eta) = p(\eta, \mu, b \mid m)$, which is the core of the FEP. In \cite{udo} this constraint is an effective potential for a non-equilibrium system, which is explored more in Section \ref{non-eq-subsection}.



The hypothesis in Corollary \ref{q-equals-p-corr} is a particular special case where the system is in perfect harmony with its environment. A consequence of this corollary is that only obtaining an identically zero minimum for \eqref{free-energy}, and the surprisal, recovers strict equality. In \cite{afepfapp}, it is stated that this constraint is not always satisfiable in a strict sense, in that $p(\eta, \mu, b \mid m)$ is not always known.\footnote{From now on we will neglect to write the model $m$, taking its existence as implicit everywhere in virtue of the system being defined.} Moreover, the attentive reader will note that our foregoing discussion (especially Definition \ref{free-energy}) defined a relative entropy functional for densities with differing supports, which formally can be nothing more than a heuristic. Observe that \eqref{free-energy} factorises into 
\begin{equation}\label{var-FE}
F(\mu, b) = \E_q\big[\ln\{q(\eta \mid \mu)\} - \ln\{p(\eta \mid \mu, b)\}\big] - \ln\{p(\mu, b)\},
\end{equation}
implying that, in fact, the free energy is the surprisal when the bound vanishes. This observation is why the FEP takes the view of bounding surprisal by minimising free energy variationally, rather than what is supposed by Corollary \ref{q-equals-p-corr}, \emph{per se}. In addition to the connection between free energy and surprisal being more evident, \eqref{var-FE} is well-defined on states $\eta$, since $\text{supp}(q) \subseteq \text{supp}(p)$ when we take parameter values as conditional events. So, for both technical and conceptual reasons, we postulate a variational approximation of $p(\eta, \mu, b \mid m)$, where the approximation is controlled by the parameter of internal states. We refer to the above functional in the ideal case as the particular free energy. For more on the distinction, see sections four and six of \cite{millidge}. 

\begin{lemma}\label{sigma-lemma}
The minimisation of free energy takes place if internal states model external states.
\end{lemma}
\begin{proof}
Assume $b$ is a Markov blanket between $\mu$ and $\eta$ such that $p(\eta \mid \mu, b) = p(\eta \mid b)$,\footnote{This is given in \cite{dacosta}.} and let $\hat \mu_b = \E_{p(\mu \mid b)}[\mu\mid b]$ (likewise for $\eta$). As stated, we also postulate that internal states parameterise external states by mapping to a sufficient statistic for $p(\eta)$ given a value of $b$, such that $p(\eta \mid b)$ is modelled by $q( \eta; \sigma(\hat\mu_b) )$ under $\sigma(\hat \mu_b) = \hat \eta_b$. This is sufficient for all of the results shown\textemdash observe that if $\sigma(\mu)$ is chosen to match the parameters of $p(\eta \mid b)$ and thus $p(\eta \mid \mu, b)$, then the KL divergence term of \eqref{var-FE} is identically zero.
\end{proof}

We also prove the following important result about internal states, elaborating on Lemma \ref{q-mu-and-b-lem} and completing Lemma \ref{sigma-lemma}. 
By proving that $\hat\mu_b$ is the minimiser of variational free energy, and bounds the surprisal, we can go on to prove the approximate Bayesian inference lemma given in \cite{afepfapp}.

\begin{lemma}\label{var-free-energy-lem}
Taking the expected internal state as the ideally free energy minimising state yields 
\[
\E_{p(\mu \mid b)}[\mu \mid b] = \hat \mu_b = \argminemph_{\mu_b}[F(\mu, b)],
\]
such that the variational free energy $\tilde F(\mu, b) = \E_q\big[\ln\{q(\eta \mid \mu)\} - \ln\{p(\eta \mid \mu, b)\}\big]$ bounds both the free energy and the surprisal from above.
\end{lemma}
\begin{proof}
Since we assume the particular free energy $F_{q^*_\mu}(\mu, b)$ reduces to the surprisal under a control parameter $\mu$, we have 
\[
\tilde F(\mu, b) \geq F(\mu, b) \geq F_{q^*_\mu}(\mu, b),
\]
which in fact follows from \eqref{var-FE}. The intended result then follows from the Bogoliubov inequality. 
\end{proof}

Lemmas \ref{sigma-lemma} and \ref{var-free-energy-lem} suggest internal states ought to be chosen such that $q(\eta \mid \mu)$ models $p(\eta \mid \mu, b)$, making $\mu$ a control parameter for the agent-environment loop (given a map $\sigma(\hat\mu_b) = \hat\eta_b)$.
The mathematical theory of variational approximation has been discussed previously by the author in \cite{mft-im}, where a proof of the role the Bogoliubov inequality plays in mean field theory appears. The above inequalities are typically strict\textemdash we do not observe identity unless the surprisal is zero, in which case $\tilde F(\mu, b) = F(\mu, b) \geq 0$, or $q$ is optimal, in which case we have the reduction mentioned before.

Importantly, Lemma \ref{var-free-energy-lem} states that\textemdash to learn the causes of internal states \emph{and} bound the surprisal of those states, it is sufficient to do variational inference. This is a key result in \cite{afepfapp}, shown there in a similar fashion. We now prove that there is an equivalent statement in the context of constrained entropy:

\begin{prop}\label{q-equals-p-var-prop}
Under the constraint that external states are on average as informative as blanket states, Corollary \ref{q-equals-p-corr} holds for \eqref{var-FE} such that $q^*_\mu(\eta) = p(\eta \mid \mu, b)$, which is approximately $p(\eta, \mu, b)$.
\end{prop}
\begin{proof}
Let $J(\eta) = -\ln\{p(\eta \mid \mu, b)\}$ and $\E_q[J(\eta)] = -\ln\{p(\mu, b)\}$, such that the expected surprisal of external states given internal and blanket states is equal to the surprisal of internal and blanket states. This yields the expression 
\[
-\int \ln\{q(\eta\mid \mu)\} q(\eta \mid \mu) \dd{\eta} - \left(- \int \ln\{p(\eta \mid \mu, b)\} q(\eta \mid \mu) \dd{\eta} + \ln\{p(\mu, b)\}\right),
\] 
recovering \eqref{var-FE} after negation. By the same argument as that of Proposition \ref{max-ent-q-prop}, this leads to maximising entropy, which yields $q^*_\mu(\eta) = p(\eta \mid \mu, b)$ as in Corollary \ref{q-equals-p-corr}.
\end{proof}

\begin{remark}\label{interpretation-rem}
If the blanket consists only of sensors, this is the constraint that external states contain no new information on average, which is equivalent to $q_\mu(\eta)$ explaining external states. Moreover, this optimal sampling is equivalent to a Markov blanket remaining in place, in virtue of self-organisation under those explanations; this is apparent in that $J(\eta)$ is also directly equivalent to a bound on the average surprisal of blanket states, and thus, implies some average structural integrity.
\end{remark}


We can now prove the approximate Bayesian inference lemma of \cite{afepfapp}. 

\begin{theorem}[approximate Bayesian inference]\label{ABIL-thm}
When conditioned on blanket states, the internal states of a self-evidencing system on average perform approximate Bayesian inference on external states, via a minimisation of variational free energy.
\end{theorem}
\begin{proof}
This follows from Lemmas \ref{sigma-lemma} and \ref{var-free-energy-lem}, in that internal states minimise free energy to reduce the average surprisal of external states to some lower bound. By Proposition \ref{q-equals-p-var-prop}, this is formally equivalent to maximising entropy, and thus is Bayesian inference over the optimal assignment of probability to any given external state. As was shown in Corollary \ref{q-equals-p-corr}, the optimal such probability is the optimal belief over states.
\end{proof}

Lemma \ref{sigma-lemma} and Proposition \ref{q-equals-p-var-prop} were particularly important for this result. We  also implicitly used the fact that maximising entropy is formally Bayesian inference against an unconstrained reference density serving as a uniform prior (see Jeffrey conditioning, for instance). We used this to suggest that the mode matching prescribed by the FEP is a maximum entropy inference of the optimal assignment of probabilities to states against some constraint. 





We now give the first major contribution of this paper, in this case, to the FEP. Previous results here established that the recognition density $q(\eta \mid \mu)$ is a maximum entropy probability density, and that all of the related FEP-theoretic statements about $q(\eta \mid \mu)$ are both sensible and have statements in the language of maximum entropy. It is initially clear that we can derive the free energy principle from maximum entropy at a conceptual level, given the fact that the free energy principle aims to produce an attractor defining a system, and claims that the system self-evidences to inhabit that region of state space \cite{cocycles-paper, maxwell2018}. Indeed, the duality between free energy over beliefs and entropy over beliefs is a bit trivial, consisting of (in Proposition \ref{max-ent-q-prop}) a negation and a compensatory maximisation. Moreover, that free energy can be derived from constraints on belief updating has been discussed in the literature before \cite{FE-from-consts}, and free energy and entropy are easily related via convex analysis.

Self-organisation frames the question in terms of self-entropy, however, and seeks to determine where such an attractor actually comes from. This perspective is not immediately apparent from \eqref{free-energy}, in that it asks about the behaviour of internal states, whilst in the FEP, the behaviour of internal states follows from an analysis of the environmental states embedding the system. Even in \eqref{var-FE}, the surprisal of internal states is only incidental to the fashioning of a model of external states. What we want is effectively to \emph{dualise} the entire construction\textemdash to ask not what external states look like from the perspective of the system, but what the system looks like from the perspective of the environment. Thus, we leverage the duality defined previously: we maintain the structure of a model of states, but it becomes a model of internal states, contingent on external states, dictating what internal states are likely given the dynamics of the system and the environment. In effect, in order to treat the system itself as more canonical, we ask about the Bayesian inversion of the recognition density, $q(\mu \mid \eta)$. 


We may sketch a proof of this equivalence as follows: assume there is a maximum entropy probability density $q(\mu)$ constrained to be localised around system-like states, or, states that are amenable to organisation according to some notion of the system. This could be a stone remaining in crystalline states, or a human remaining in allostatic states. The presence of blanket states, being a proxy for surprisal, implies that to every external state and internal state is associated a blanket state, such that external states cause blanket states, which are paired at any time with possible internal states $q(\mu \mid \eta)$. From there, the pairing of highly likely internal states with low surprisal blanket states is equivalent to the FEP. 
Conceptually, this follows by tautology: if a state $\mu$ satisfies $\sigma^{-1}(\hat\eta_b)$, it models a density over external states, meaning it is a state that makes organisation possible, and thus is system-like. If a state is system-like, it draws from a system-like density over states. The other direction of implication easily follows from this argument, since minimising surprisal on external causes of changes in internal states maintains a locus of system-like states. Explicitly, we have the following technical lemmas (elaborations on results found in \cite{dacosta}; see there for a relevant proof and expository material) and theorem:

\begin{defn}
Let $[-,-]$ be the set of all maps between two objects, and let there be an evaluation map $\emph{ev}$ taking a mapping set and evaluating a map therein on an input tensored into the arguments of $\emph{ev}$, e.g., $\emph{ev}([X, Y] \otimes X) \to Y$ by taking $f : X \to Y$ and passing into it some $x \in X$, yielding $f(x) = y \in Y$. The evaluation of maps to produce objects is adjoint to the composition of objects, which produces maps. Thus, given appropriate symmetry assumptions, one exists if and only if the other does.
\end{defn}

Recall that a synchronisation map $\sigma$ exists, taking the expected internal state given a blanket state to the corresponding expected internal state for that blanket state. We prove some qualities of the synchronisation map now; we do so in category-theoretic language following our description in Section \ref{duals-subsection} of the categorical structure of synchronisation.

\begin{lemma}\label{tensor-hom-lem}
The synchronisation map $\sigma$ exists injectively if and only if the factorisation $\sigma = \bm\eta\circ\bm\mu^{-1}$ exists and $\bm\mu$ and $\bm\eta$ are injections. 
\end{lemma}
\begin{proof}
We have constructed $\sigma$ such that it sends $\hat\mu_b$ to $\hat\eta_b$. To define such a function we take the evaluation map of $[\hat\mu_b, b]$ and a fixed $\hat\mu_b$, getting $b$ out of it; we follow this by feeding that $b$ to $[b, \hat\eta_b]$. Overall we have
\[
[b, \hat\eta_b] \otimes ([\hat\mu_b, b] \otimes \hat\mu_b) \to [b, \hat\eta_b] \otimes b \to \hat\eta_b,
\]
which accepts and returns the same quantities as $\sigma$. Applying tensor-hom adjunction, this relation is equivalent to the composition of two maps, 
\[
[b, \hat\eta_b] \otimes [\hat\mu_b, b] \to [\hat\mu_b, \hat\eta_b].
\]
Hence, $\sigma$ is equivalent to the factorisation indicated, proving one part of the claim. Now, restricting an injection to its image set produces a bijection, and the composition of two bijections is again a bijection. Since this factorisation is possible, clearly $\sigma$ is injective if and only if both $\bm\eta$ and $\bm\mu$.
\end{proof}

In effect, Lemma \ref{tensor-hom-lem} allows us to claim that $\sigma$ assigns each blanket state to a map between an expected internal and expected external state. Since $\sigma = \bm\eta \circ \bm\mu^{-1}$, these two sets are identical. 
It is important to note that, due to the compositional structure of $\sigma$, it is \emph{not} explicitly a function of blanket states. Only internally (i.e., within the sequence of maps) does the blanket state matched to the expected internal state get changed into the expected external state given that blanket state. 

To make the above statement clear, we can expand the adjunction in Lemma \ref{tensor-hom-lem} to
\[
[ [b, \hat\eta_b] \otimes [\hat\mu_b, b], [\hat\mu_b, \hat\eta_b] ] \cong [ [b, \hat\eta_b], [[\hat\mu_b, b], [\hat\mu_b, \hat\eta_b]] ],
\]
which gives $\bm\eta \circ \bm\mu^{-1} = \sigma$ on the left and reconstructs $\sigma$ equivalently on the right as a function which generates $\hat\eta_b$ out of $(\hat\mu_b, b)$ by an internal function $\bm\eta : b \to \hat\eta_b$. In other words, $\bm\eta \mapsto (\bm\mu^{-1} \mapsto \sigma)$. If we so choose, we can assign a fixed $b$ to $\sigma$, instead of producing it internally from evaluating $\bm\mu^{-1}$ as we did above. This changes the definition of $\sigma$, but interestingly, it produces the same result. The same adjunction on this function, $[b \otimes \hat\mu_b, \hat\eta_b] \cong [b , [\hat\mu_b, \hat\eta_b]]$, constructs a function of pairs, $\xi(b, \hat\mu_b)$; for an unknown secondary input $\hat\mu_b$, fixing\footnote{Evaluating a function of products on a single input is referred to as \emph{currying} in type theory and parts of functional analysis. The author thanks Dan Abramov for this reference.} $b$ produces $\sigma$:
\[
\bm b : b \mapsto \xi(b, -),  \quad \xi(b, \hat\mu_b) = \sigma(\hat\mu_b) = \hat\eta_b.
\]
This obscures the compositional structure of $\sigma$, however, making it of little use to us. The shared structure of functions of opposite inputs under the tensor product is a reflection of braiding. Indeed, by checking braiding relationships, then reapplying tensor-hom adjunction and de-evaluating $\hat\mu_b$ and $\hat\eta_b$, it is straightforward to prove that these definitions of $\sigma$ imply each other. 


The non-trivial structure of $\sigma$ can be seen as a symptom of the highly structured relationship encoded by $\sigma$, namely, the assignment of external states to internal states via an assignment of identical blanket states to both.

\begin{lemma}\label{marginal-sigma-map-lem}
There exists a map taking the marginally expected internal state $\hat\mu$ to the marginally expected external state $\hat\eta$. Moreover, this map is equivalent to $\sigma(\hat\mu) = \hat\eta$.
\end{lemma}
\begin{proof}
Let $\bm\mu : b \mapsto \hat\mu_b$ and $\bm\eta : b \mapsto \hat\eta_b$ injectively, and assume $\sigma(\bm\mu(b)) = \bm\eta(b)$, factorising into $\sigma : \hat\mu_b \xrightarrow{\bm\mu^{-1}} b \xrightarrow{\bm\eta} \hat\eta_b$. By the partition theorem, 
\[
\E_{p(\mu)}[\mu] = \E_{p(b)}[\E_{p(\mu\mid b)}[\mu \mid b]],
\]
which we have denoted as $\hat \mu$ and $\E_{p(b)}[\hat\mu_b]$, respectively (and likewise for $\hat\eta$ and $\E_{p(b)}[\hat\eta_b]$). Thus, we seek to construct a map 
\[
\tilde\sigma( \E_{p(b)}[\E_{p(\mu\mid b)}[(\mu \mid b)]]) = \E_{p(b)}[\E_{p(\eta\mid b)}(\eta\mid b)].
\]
By construction, $\bm\eta$ and $\bm\mu$ are linear functions of blanket states, and since the composition of linear functions is again linear, $\sigma$ is linear in $b$. As such, we have the following identity, which holds in the sense of the Bochner integral:
\[
\E_{p(b)}[\sigma(\hat\mu_b)] = \sigma(\E_{p(b)}[\hat\mu_b]),
\]
implying that $\E_{p(b)}[\sigma(\hat\mu_b)] = \E_{p(b)}[\hat\eta_b]$ if and only if $\sigma(\E_{p(b)}[\hat\mu_b]) = \E_{p(b)}[\hat\eta_b].$
\end{proof}

If the fairly reasonable assumptions of the above lemma are not met\textemdash say, if $\mu$ is not Bochner integrable as a random variable, or $\sigma$ is not continuous\textemdash we can still construct a marginal synchronisation function $\tilde\sigma : \hat\mu \mapsto \hat\eta$ quite easily. Lemma \ref{marginal-sigma-map-lem} is simply an ideal case where our existing $\sigma$ does the work for us. Establishing as much is an important technical step for the proof of Theorem \ref{max-ent-FEP-thm}, which proves that systems which perform approximate Bayesian inference under a Markov blanket maximise constrained self-entropy, in line with Claim \ref{FEP-is-max-ent-claim}. We first prove the given statement for a fixed blanket state, and then prove it marginally, such that each $\hat\mu_b$ samples from some system-like $q(\mu)$; invoking $\tilde\sigma$ is imperative to this notion of system-ness.


\begin{theorem}\label{max-ent-FEP-thm}
Given a fixed set of constraints, the probability measure in the state space is the constrained maximum entropy probability distribution on internal states if and only if 
the surprisal term in \eqref{var-FE} is on average equal to $\sigma^{-1}(\hat\eta)$.
\end{theorem}
\begin{proof}
We begin by defining a target density over states $p(\mu \mid \eta, b) = p(\mu\mid b)$ and a current density over states $q(\mu \mid \eta_b)$, taking roles dual to the optimal belief and the recognition density, respectively. We moreover we suppose we have defined a stationary sufficient statistic for $p(\mu \mid \eta, b)$, given by $\hat\mu_b$. We can construct a dualised maximum entropy density constrained such that $\E_{p(\mu \mid b)}[\mu \mid b] = \sigma^{-1}(\hat\eta_b)$, 
\[
q^*(\mu \mid \eta_b) = \argmax\left[S[q(\mu \mid \eta)] - \lambda\left(\E_{p(\mu \mid b)}[\mu \mid b] - \sigma^{-1}(\hat\eta_b)\right) \right],
\]
which exists if and only if $q^*(\mu \mid \eta_b) = p(\mu \mid b)$. To show that this implies $-\ln\{p(\mu, b)\}$ equals $\sigma^{-1}(\hat\eta_b)$ on average, note we have 
\[
-\ln\{p(\mu, b)\} = -\ln\{p(\mu \mid b)\delta(b-\bar{b})\} = -\ln\{p(\mu \mid \bar{b})\}
\]
for a given blanket event $\bar{b}$. The desired result is now a consequence of an exponential distribution having its mean as its sufficient statistic, i.e., that 
\[
\E_{p(\mu \mid b)}[-\ln\{p(\mu \mid b)\}] = \E_{p(\mu \mid b)}[\mu \mid b] = \sigma^{-1}(\hat\eta_b).
\]
This produces one direction of the dual result to the approximate Bayesian inference lemma (Theorem \ref{ABIL-thm}, here)\textemdash that if we maximise self-entropy, we bound surprisal, and perform inference over what the optimal $q(\mu \mid \eta)$ is. The other direction of implication follows easily from the converse of the claim: if $p(\mu\mid b)$ satisfies $\E_{p(\mu\mid b)}[-\ln\{p(\mu\mid b)\}] = \sigma^{-1}(\hat\eta_b)$, then by definition of a sufficient statistic, it is automatically the maximum entropy probability density given that constraint. Finally, using Lemma \ref{marginal-sigma-map-lem}, we can deduce the marginal result
\begin{equation}\label{thm-2-main-eq}
q^*(\mu \mid \eta) = \argmax[S[q(\mu \mid \eta)] - \lambda\left(\E_{p(\mu)}[\mu] - \sigma^{-1}(\hat\eta)\right).
\end{equation}
This admits the same argument as the first portion of the proof, in virtue of the more general fact that total conditionalisation and marginalisation preserve identities.
\end{proof}

A slightly weaker statement than that of Theorem \ref{max-ent-FEP-thm} is that an upper bound for surprisal exists, in which case, the maximum entropy density under the constraint given is not necessarily the best description of the system; the total constraint on any given state may be greater or less than what is required. An alternative proof of Theorem \ref{max-ent-FEP-thm} constructs a sampling process that converges in law to $q(\mu \mid \eta)$; to see why, and to deconstruct the technical portions of the above theorem, we can revisit the example of the stone:
\begin{example}
To model a stone on a beach, we might choose to model how the temperature changes in the environment affect internal states via a particular blanket state. During the day, when the blanket is in a state of receiving the sun's rays, the expected temperature at a point on the interior of the stone is given by whatever function corresponds to the heat flow into that material given the intensity of sunlight on its boundary. Conversely, at night, the blanket state is no longer that of `receiving sunlight,' and so the internal temperature of the stone at a point is given by the diffusion of heat out of the stone given the coolness of the environment surrounding it\textemdash say, the temperature and speed of the land breeze making contact with the stone. In neither case is the stone entering non-stone-like states. In both cases, the stone is (by construction) occupying the expected state for that blanket state on average. Now, suppose the day time heat becomes so intense the stone melts. That the stone still goes to the expected internal state given the blanket state is the tautology of evidencing\textemdash but, the fact that the expected internal state is no longer stone-like is a result of the blanket being passed down to a different set of states, rather than the initial set of stone-like blanket states. That is to say, suddenly, blanket states are surprising. In this case, ours is now a poor model of the stone. Clearly that is because it is no longer stone-like, but also, because we fail to meet the condition that $\E_{p(\mu \mid b)}[-\ln\{p(\mu , b)\}] = \sigma^{-1}(\hat\eta_b)$ for that $b$; in fact, the surprisal of $b$ (here, joint with $\mu$) is much greater. This reflects the fact that our old Markov blanket does not exist anymore, and has moved to a new regime of states. Correspondingly, a new model for which those states are not surprising is possible, but requires re-calibrating the probabilities of blanket states.
\end{example}

When we discuss attractor states for stones in Section \ref{sample-paths-sec}, we give another useful example (Example \ref{stone-attractor-example}).

These results clarify the intuitively un-obvious statement of Claim \ref{FEP-is-max-ent-claim}\textemdash that minimising free energy over beliefs maximises constrained self-entropy. This is crucial to relating the FEP to self-organisation. In fact, we do this by focussing on the surprisal, rather than re-dualising the construction, which would have been completely uninsightful. It is fitting that the FEP also focusses on its variational case, drawing from insights about the surprisal of internal states, to comment on self-organisation.

Concomitantly, these results can be informally stated as: for a system which is optimally system-like, external interventions\textemdash and thus their observations\textemdash cannot be surprising. If they are, then this enacts changes to unpreferable states, such that the system fails to remain organised\textemdash or, in the adaptive case, must act to re-organise. If they are not, states being expected means the system draws from the maximum entropy $p(\mu \mid \eta)$ under $\sigma$.\footnote{Accordingly, note how Theorem \ref{max-ent-FEP-thm}\textemdash as well as its dual on $q_\mu(\eta)$\textemdash clearly both fail if $\sigma$ is not injective. This is an elementary property of functions which are left-cancellative, which we require by taking $\sigma^{-1}$; it has also been noted in \cite{parr, dacosta, aguilera2021}.} As such, the maximum entropy view follows from both the interpretation and definition of the FEP. 
In theory, it is now possible to\textemdash without appealing to external conceptualisations of what is system-like about a system\textemdash summarise a sufficient set of constraints as the set of constraints that makes internal states unsurprising, according to the above identification. This returns us to the idea of existential variables discussed in Section \ref{constraints-as-blankets} and \cite{Andrews2020}. 

\begin{remark}
Importantly, we could have constrained \eqref{thm-2-main-eq} and any of the other results here by the \emph{distance} of a state away from the intended mean, as in Remark \ref{variance-constraints-remark}. This would introduce a Laplace approximation to the true density $p$, which\textemdash fittingly\textemdash appears in \cite{dacosta} after introducing additional covariance information to $\sigma$. See Theorem \ref{eff-eq-thm} for more on how this is implicated in non-equilibrium steady states. This additional constraint is often important for equilibria alike, in that it contains information about the physics of a situation\textemdash this can help clarify the FEP's role in modelling specific systems, which is broadly separate from its mathematical character. Again, we speak more about this later, in Section \ref{non-eq-subsection} especially.
\end{remark}

We end the beginning of Section \ref{max-ent-FEP-section} by discussing one further idea. In the following corollary, we reference a result due to \cite{pinkus}\textemdash namely, that a non-polynomial function of a polynomial can approximate any continuous scalar function arbitrarily well (see Theorem 3.1 there).

\begin{corollary}
Maximum entropy can approximate any process to arbitrary precision under FEP-theoretic constraints.
\end{corollary}
\begin{proof}
Let $\mathscr{C}^0(\R^n; \R)$ be the space of $\R$-valued continuous functions on an $n$-dimensional real domain. Suppose $J(\mu)$ can be expanded into a power series of constraints on acceptable states. The exponentiation of such polynomials is dense in $\mathscr{C}^0(\R^n; \R)$. Thus, it is a universal approximator for the classification of states of $\mu$-valued systems when $\mu \in \R^n$. Classifying $\mu$ by the probability of observing $\mu$ solves the Fokker-Planck equation describing $p(\mu)$.
\end{proof}

The above corollary recalls the architecture of a neural network by showing that our model of FEP-type systems is capable of computing the probability density over states of the system to arbitrarily small error. In translating from formal ideas to physical ideas, it is important\textemdash it implies there are situations where the constraint viewpoint is only precise if we have arbitrarily detailed information about the system. The tractability of constraints in practice thus depends highly on how complicated such a system is.

\subsection{Geometry and analysis}\label{geometry-and-analysis-sect}

This completes the first set of results in this paper. Having established that the FEP is equivalent to the principle of maximum entropy, which puts Bayesian mechanics in the language of probability theory and stochastic dynamical systems theory, we will now discuss the accompanying geometric and analytic theory available to us. There is a sense in which the FEP could only work if there existed some deeper connection between the gradient flows of Lagrangian mechanics and the information geometries of statistical mechanics, establishing both a geometry and functional calculus for Bayesian mechanics. The following material relies on some obvious, but largely un-investigated, connections between information geometry in stochastic dynamical systems and symplectic geometry in classical dynamical systems. We will build up the theory as necessary, but leave a complete account of the symplectic geometry of statistical physics to future work.

The following theorem is the first significant result in this portion of Section \ref{max-ent-FEP-section}. The statement is intuitively obvious, owing to the shared roles of potential and constraint functions in shaping the dynamics of a process. In the simple case of a Wiener process, we have a set of useful formal results from the analysis of functionals and partial differential equations. We will make particular use of results due originally to Bakry and \'Emery, and Markowich and Villani. 
Many of these results use the minimisation of the positive entropy rather than the maximisation of Shannon entropy, for technical reasons; clearly, these are equivalent. In fact, we used this observation to prove Proposition \ref{max-ent-q-prop} and other results earlier in this section. In that spirit, all of these statements hold after adjunction, to describe a maximum entropy model of external states as well. Due to this symmetry, we typically need not specify variables of interest; as such, to follow references \cite{villani} and \cite{ledoux}, we will switch to the functional notation $\int p \ln p$ for entropy. 

\begin{theorem}\label{constraint-is-potential-thm}
In the equilibrium case of a Wiener process, the constraint on a maximum entropy problem is a potential function for the dynamics of the underlying sampling process.
\end{theorem}
\begin{proof}
Suppose the noise is a Weiner process with the diffusion matrix set to the identity, e.g., $\gamma_t$ is given by 
\[
\dd{\gamma_t} = - \grad V(\gamma_t) \dd{t} + \dd{W_t}.
\]
The associated Fokker-Planck equation 
\begin{equation}\label{fokker-planck}
\pdv{p}{t} = \Delta p  - \grad V \cdot \grad p
\end{equation}
converges exponentially quickly to the stationary solution $\exp\{- V \}.$ Taking the gradient descent of the functional 
\[
\int p \ln p - \lambda \int J p + \lambda C
\]
yields 
\[
\int p \left( \partial_t p + \grad \cdot (p \lambda J) - \Delta p \right) = 0
\]
after integrating by parts, which is a Fokker-Planck equation with $\lambda J(x) = V(x)$.
\end{proof}

In effect, we have proven that the diffusion process which maximises entropy diffuses in a potential given by the constraints. Intuitively this is rather obvious, since the potential function changes the shape of the gradient of the entropy. 

We proceed with this set of analytic results by demonstrating a related idea from the functional analysis of entropy. First, we prove a modified logarithmic Sobolev inequality. See \cite{ledoux} for a review of the actual log-Sobolev inequality.

\begin{lemma}\label{mod-log-sob-lem}
For a process sampling from a measure of Gibbs-type, the classical kinetic action of its density dynamics bounds the positive entropy of its density from above.
\end{lemma}
\begin{proof}
Let a measure of Gibbs-type be a probability density $p = Z \exp{-V}$ for some potential $V$ which is positive 
on the support of $p$, and a fractional constant $Z$. In other words, a maximum entropy probability density under a constraint function $V$. Since, by physical arguments, the kinetic energy of a system is always positive, the kinetic action\textemdash an integral of a positive quantity\textemdash is also always positive; in contrast, the Shannon entropy of a process is generally positive, such that the positive entropy is negative. This inspires the \emph{ansatz} 
\[
\int p \ln p \dd{x} \leq \frac{1}{2} \int \abs{\partial_t p} ^2 \dd{t}.
\]
For a Gibbs-type measure, the positive entropy evaluates to:
\[
\int Ze^{-V} \ln Ze^{-V} \dd{x} = -\int Ze^{-V} \left(\ln\frac{1}{Z} + V\right) \dd{x}.
\]
Let $Z = k^{-1}$ for some $k > 1$. Since $Z\exp{-V} > 0$ and $\ln k + V \geq 0$, the overall integral is negative. Now, since $\abs{ \partial_t p}^2$ is always positive, we have the relation
\[
\int p \ln p \dd{x} \leq 0 \leq C \int \abs{\partial_t p} ^2 \dd{t}
\]
for any $C \geq 0$, which proves the claim.
\end{proof}

We make no undue claim to novelty by providing Lemma \ref{mod-log-sob-lem}, since this result offers no greater analytic insight than the true log-Sobolev inequality, beyond implying it whenever $\abs{\partial_t p}^2 \leq \grad p \cdot \grad p$. What thus \emph{is} insightful is that we can deduce the log-Sobolev inequality purely by physical arguments about the behaviour of the entropy functional.

\begin{corollary}\label{diff-min-cor}
If a diffusion process minimises its classical kinetic action, then it bounds its free energy from above.
\end{corollary}
\begin{proof}
Let $\partial_t p = 0$. Since the classical kinetic action is quadratic in velocities, this is an absolute minimum for the upper bound in Lemma \ref{mod-log-sob-lem}. Thus, enforcing a stationary probability density bounds the positive entropy from above, which is a lower bound on the negative entropy; by Theorem \ref{max-ent-FEP-thm}, this places an upper bound on the free energy. In particular, evaluating $\int p \ln p$ for a Gibbs-type measure yields $- \int V p$ which is a constraint on a maximum entropy problem.
\end{proof}


This proves one direction of the particular free energy lemma that FEP-type systems are least action systems. The result recalls that of Lemma \ref{stationarity-steady-state-lem}\textemdash the model of thing-ness provided by the FEP is valid for any `thing,' given a definition of a thing as an object with some key invariant characteristics. We take the position that this generality is both a boon and a bane. The FEP is a powerful modelling framework, but (as written here) has nothing special to say about any particular system being modelled. See also remarks in Section \ref{as-if-inference-subsec}.


Using some well-known identities relating to the entropy functional, all of which are noted or proven in Markowich and Villani's seminal paper \cite{villani}, we can also prove that maximising entropy under the above equivalences induces an information geometry. Critically, the following proposition says very little of sample paths or dynamics on that statistical manifold, only proving that it exists asymptotically.

\begin{prop} 
Stationary points of the free energy functional are minima of the Fisher information.
\end{prop}
\begin{proof}
For any convex function $\psi : \R \to \R$ and $p$ satisfying \eqref{fokker-planck}, the functional 
\[
\int \psi(p) p 
\]
is a Lyapunov function for \eqref{fokker-planck}. Choosing $\psi(p) = \ln p  - V,$ we have both constrained positive entropy and free energy, \begin{equation}\label{free-energy-lyapunov-eq}
\int \ln\{p\} p - \int V p.
\end{equation}
This functional metrises a space of probability measures by the Csisz\'ar-Kullback inequality. Moreover, the induced flow of $p$ by this Lyapunov function obeys the Fisher information, since 
\[
\dv{t}\left( \int p \ln p - \int V p \right) = \dv{t} \int \ln\left\{\frac{p}{e^{-V}}\right\}p = -\int \abs{\grad\ln\left\{\frac{p}{e^{-V}}\right\}}^2 p.
\] 
When $p = \exp{-V},$ for appropriate boundary conditions, both quantities are zero.
\end{proof}

Note that the Fisher information is equivalent to $\int \abs{\grad \sqrt{p} }^2 $, calculated by properties of the logarithm. This provides us with an equivalent formulation of the actual log-Sobolev inequality \cite{ledoux}. As such, this also proves a weak formulation of the other direction of the particular free energy lemma, by establishing an equivalence between least action and free energy.

We can give some further geometric results. The information geometric aspect of the flow of $p$ is an important property of both the FEP and the maximisation of entropy, especially when we discuss the sort of iterated inference that characterises non-stationary systems. In maximising entropy, we have a (potentially small) number of system-like states occupied with high probability. 
A system which is ideally itself is ideal with respect to the set of constraints that characterises that system. This means that if a system fails to self-evidence, it may appear to obey a different set of constraints. The correspondence allows us to prove another theorem:



\begin{theorem}\label{change-in-J-info-geo-thm} 
Determining the optimal $p(\mu; J')$ after a change in constraints $J \mapsto J'$ induces an information geometry.
\end{theorem}
\begin{proof}
Assume the KL divergence can be made symmetric, for instance, by taking its second-order Taylor expansion, and that $p(\mu; J)$ and $p(\mu; J')$ have the same support. We show that the stochastic channel created by changing a set of constraints metrises a space of probability densities, thus constructing a moduli space $\mathcal{P}$ containing points $p_J$ with distances $D_\text{KL}(p_J, p_{J'})$. Suppose the system has already optimised for $J$ such that its generative model is $p(\mu; J)$. Under $J \mapsto J'$, we can combine the two actions and define an invariant measure such that 
\[
S[p;J'] = \int \ln\left\{\frac{p(\mu; J')}{p(\mu; J)}\right\} p(\mu; J') \dd{\mu}. 
\]
This is the relative entropy, which, upon maximising its negation, minimises the KL 
divergence between the current $p(\mu; J)$ and the target $p(\mu; J')$.
\end{proof} 

The consequence of this theorem is that adaptive self-evidencing is a movement in `thing-ness' space, wherein a system must minimise the distance between one density parameterised by one set of $J$ and the intended density parameterised by $J'$. This is done by re-maximising entropy against the target set of constraints.

\begin{corollary}
Failure to self-evidence according to one set of constraints leads to a new set of constraints prescribing the optimality of the resultant structure. 
\end{corollary}
\begin{proof}
This follows from Theorems \ref{max-ent-FEP-thm} and \ref{change-in-J-info-geo-thm}.
\end{proof}

This perspective formalises arguments in the literature commonly referred to as the `passing down' of Markov blankets. Refer to Example \ref{inertness-example} to see how this describes a phase transition in the possible dynamics of the system. In humans, this is the transition from life-like states to a structure which is inert, after drifting too far away from the allostatic set-points dictated by $J$. Such a structure is still organised, and thus still evidences according to some new set of constraints $J'$. 

At this point we pause and defer 
any further geometric ideas, whilst we take an extended detour into the geometry and physics of quantum field theory in Sections \ref{max-ent-field-section} and \ref{gauge-symmetry-section}. We will take up the FEP again later in Section \ref{gauge-symmetry-section}. Before that, we take the opportunity to address some measure-theoretic questions from the analysis of random variables, as well as a philosophical point about these results.

\subsection{Technical remarks on random dynamics}\label{technical-remarks}


Some apparent technical issues arising in \cite{afepfapp} are easily patched in light of these results, as maximum entropy has strong connections to formal probability theory and the analysis of random functions; additionally, we can begin to give an explicit characterisation of the attractor of system-like states we are interested in. One key problem in this context is that of ergodicity assumptions in the FEP. In the following results, we assume the state space has a $\sigma$-algebra\footnote{To avoid any confusion, we point out that this has nothing to do with the synchronisation map of Sections \ref{bayes-mech-prelim-sect} and \ref{max-ent-FEP-section}, and is so named as a convention within probability theory.} of open sets $\F$, preferably the set of Borel sets admitted by $X$, collecting states as observable events. Moreover, we assume $(X, \F, P)$ is a complete measure space or extends to one. Throughout, $x$ is an arbitrary state variable. We begin with the following corollary: 
\begin{corollary}\label{stationary-process-cor}
It is necessary that the target density $p(x)$ be stationary for particular free energy to be minimised.  
\end{corollary}
\begin{proof}
Observe that the maximum entropy Lagrangian has no velocity coordinate, and thus assumes that $\partial_t p(x, t) = 0$ on any 
level set of $S[p]$. Moreover, the constraining environmental density is necessarily fixed for fixed constraints.
\end{proof}

This affirms the equivalence of free energy minimising systems and systems at steady state given in Lemma \ref{stationarity-steady-state-lem}, clarifying the role of fixed constraints in defining system-ness.

\begin{defn}\label{ergodic-def}
Let $A \in \F$. An ergodic probability measure is a probability measure for which no subset $P(A)$ is invariant under a measure-preserving function $T : X \to X$; that is, $P(T^{-1}(A)) = P(A)$ and $P(A) = 1$ for all $A$ up to sets of measure zero.
\end{defn}

Taking $T$ as a shift map inducing the flow of a random dynamical system \cite{ludwig}, Definition \ref{ergodic-def} is intuitively that `every meaningful state will certainly be visited by the system'\textemdash that is, system-like states are observed almost surely. Based on this, we introduce a notion of \emph{local ergodicity}, described below. The idea of local ergodicity is not completely vacuous, in that states of probability measure zero are not impossible\textemdash thus, they are contained in the state space\textemdash but there is not a meaningful sense in which the system occupies those states, which is certainly (tautologically) true if they are not system-like. Taking the Bayesian interpretation of probability, our model of the system places no belief in non-system-like states. Moreover, an event which is possible but has measure zero should be surrounded by other possible events, but we demand that an entire open set around any such event has measure zero, which is a slightly stronger claim. 

Since $\sigma$-algebras are closed under complement and intersection, locally ergodic measures may be constructed for systems that are ergodic on a subset of states.

\begin{defn}\label{local-ergo-def}
A locally ergodic system is a system that possesses an invariant probability measure on a subset of the state space, such that the system occupies a distinct open subset of states almost surely.
\end{defn}

The following is a minor technical lemma, a shortcut to the disintegration theorem:

\begin{lemma}\label{lariat-lem}
The retraction of measurable sets $\ell(C) = A$ is such that a measurable function $f$ on $X$ restricts to $A$ under $f(\ell(C)) = f_A(C) = f(A)$. Moreover, $\F$ is closed under pullback by $\ell$.
\end{lemma}
\begin{proof}
Let $C$ and $A$ be sets in $\F$ with $A \subseteq C$, and $\ell$ a map retracting $C$ to $A$ by the left-inverse of the inclusion $\iota : A \hookrightarrow C$. Pulling back $f$ by $\ell$ defines $f(A)$, the restriction of $f$ to $A$. The algebra $\F$ is, by definition, closed under complement. Thus we can topologise any retraction as a subset of $X$.
\end{proof}

We refer to the map $\ell$ as a \emph{lariat}. Its intuitive role in the following proposition is that of a rope tightening around system-like states; it will allow us to construct a locally ergodic measure explicitly.

\begin{theorem}\label{ergodic-thm}
A system minimises its free energy if and only if it is locally ergodic, given that it visits all system-like states.
\end{theorem}
\begin{proof}
Using the convex duality of constraints on states in \eqref{max-ent-action}, we strengthen the constraints on the subset of system-like states $A$ until non-system-like states comprise a subset of measure zero under $\int_A p(x) \dd{x} = P(A)$ for $x\in A$. Assume that $A$ occurs almost surely\textemdash that every system-like state is visited eventually. By Lemma \ref{lariat-lem}, it follows that $P(A)$ is an ergodic target distribution under $P(A) = 1$ and $P( \F \setminus A ) = 0$, and under Definition \ref{local-ergo-def}, it is a locally ergodic measure on $A$. By Theorem \ref{max-ent-FEP-thm}, the existence of $P(A)$ is equivalent to minimising free energy. By Theorem \ref{change-in-J-info-geo-thm}, the logical inverse also holds. 
\end{proof} 

Note that, since we have assumed we are working in a complete measure space, every subset of $\F \setminus A$ is also a set of measure zero. Never do we show that subsets of measurable sets are measurable: this is clearly false in general. 

\begin{remark}
Much like ergodicity itself, this can be stated heuristically as `system-like states are certain and non-system-like states are meaningless given the system is a system,' which is precisely the idea of self-evidencing. In the highly determined case, the trivial ergodic measure on a fixed point $c$ is often denoted $\delta_c$. Such a measure can be constructed under the retraction of $C$ to the point, as in Lemma \ref{lariat-lem}. We note that the local ergodicity in Theorem \ref{ergodic-thm} is more properly thought of as a definition of system-ness, since enforcing an identically zero measure on non-system-like states is technically valid under $\ell$'s re-topologisation of $X$, but is not very instructive, and fails if the constraints cannot be strengthened as required in Lemma \ref{lariat-lem}. Thus, local ergodicity is equivalent to a well-defined system-ness. The definition of local ergodicity as ergodicity on system-like states is consistent with current FEP literature \cite{maxwell2018}.
\end{remark}



To review these results: a consequence of the Corollary \ref{stationary-process-cor} is that we require the statistics of any such process to be time-wise equivalent after a single maximisation in order for the derived $p^*(x)$ to remain valid\textemdash this is intuitively sensible, but it poses a challenge to generalising the FEP to non-stationary environments. The inclusion of certain complex systems possessing learning or memory forces us into \emph{ad hoc} methods, such as the iteration of inference, or the use of effective equilibrium for stationarity (cf. Remark \ref{de-Finetti-rem}); however, the sought generalisation should be possible in the future. Non-ergodicity appears to be a non-issue given we accept stationarity, due to the lariat argument in Theorem \ref{ergodic-thm}. The central tautology at the heart of the FEP\textemdash that things which exist, exist\textemdash has already been discussed as a consequence of Lemma \ref{stationarity-steady-state-lem}. 
Consistent with the consequences of that lemma, maximum entropy makes no detailed balance assumptions\textemdash detailed balance must be specified as a structural constraint on states in maximum entropy \cite{detailed-balance}, and so \emph{a priori} nothing is said of the requirement or dismissal of detailed balance. However, note that in Remark \ref{limitations-rem} we will discuss how \emph{local} detailed balance is indeed required to model any such system.

An interesting physical consequence of these results is that adaptive systems cannot be said to violate the second law of thermodynamics, $\partial_t S[p(x)] \geq 0$.
\begin{remark}[the second law of thermodynamics]\label{2nd-law-rem}
A consequence of Proposition \ref{max-ent-q-prop}, reflected in the maximisation of the entropy of a density over beliefs about external states, is that adaptive systems do not avoid the second law of thermodynamics; rather, they leverage it, offloading the increase in entropy to their environments, and changing their beliefs accordingly. Much like self-evidencing is native to the constrained entropic view but is still apparent in the free energy view, this disordering is a signature of free energy present on the constraint-based side of the adjunction, in that this is what creates the aforementioned ontological potential which organises the system into itself. It is thus the case that adaptive systems are engines which `eat' order and produce entropy, disordering their environments to keep themselves organised.  It has indeed been argued that complex systems are in fact statistically favoured due to their role as vehicles of dissipation \cite{kai, kai2}. An example of this dissipation is seen in the conversion of edible substances to heat, such as what occurs in human metabolism. Dually, Theorem \ref{max-ent-FEP-thm} states that organised systems do indeed maximise self-entropy over system-like states, accepting the second law up to what is allowable within the confines of system-ness. This lack of determination of what constitutes a system-like state is important for the flexibility and itinerancy characterising adaptive systems, and more generally, models an inexorable tendency for agents to fluctuate and explore.
\end{remark}

In fact, a stronger statement is possible, based on this duality. We have established that an agent increases the entropy of its beliefs to within some constraints on what is an accurate belief. Non-agentially, the statistical structure of an object reflects the ever increasing entropy of its surroundings, under the constraint that this environment must still allow that object to exist (Proposition \ref{q-equals-p-var-prop}). An interesting feature of adjunctions is that there are typically signatures of either view on each side\textemdash for instance, the constraints on surprisal that appear in both Lemma \ref{var-free-energy-lem} and Theorem \ref{max-ent-FEP-thm}. The accuracy constraint of Proposition \ref{q-equals-p-var-prop} is a signature of the maximum self-entropy view on the side of beliefs\textemdash recall, adjointly, we have established that we can model an object by maximising the entropy of our own beliefs about the system, assuming the system can be understood as maximising its entropy against some constraints on what it means to be such a system. In both views we suggest a belief corresponds to a physical maximisation of entropy, in that the believed and actual probabilities of states disperses. For instance, in Remark \ref{2nd-law-rem}, we suggest this belief includes or reflects the process of environmental disorder. 

Separately from the above results, it has been shown that correlations between coupled subsystem can (and should) be incorporated as a structural constraint on maximum entropy, leading to subadditivity \cite{dill-non-add}. This motivates the following result, again elaborating on the role of constrained maximum entropy in self-organisation:

\begin{theorem}\label{joint-entropy-thm}
Under the adjunction defined in Theorem \ref{max-ent-FEP-thm}, the joint entropy of the agent-environment loop is bounded by the total entropy of the agent-environment loop.
\end{theorem}
\begin{proof} 
Under the coupling induced by the Markov blanket, the mutual information 
\[
I[\mu ; \eta] = S[\mu] + S[\eta] - S[\mu,\eta] > 0
\]
such that joint agent-environment entropy is subadditive,
\[
S[\mu, \eta] < S[\mu] + S[\eta].
\]
When both systems maximise their entropies subject to a constraint that implies the presence of a Markov blanket, 
the total entropy of the joint agent-environment system should be bounded strictly by the sum of the entropies for each individual system, such that introducing constraints which imply a coupling decreases the joint entropy.
\end{proof}

Since the entropy of the agent-environment loop is controlled by the individual components of the loop, and introducing constraints that imply a coupling decreases that joint entropy, the integrity of the loop is precisely dependent on the integrity of its components. This statement is in some sense a statement that, by maximising constrained entropy both ways\textemdash or identically, by leveraging the coupling between agent and environment\textemdash self-organisation is possible. In so doing, it reveals the importance of the Markov blanket formulation, as a symmetric statistical relationship between the system and its environment. Likewise, this joint entropy decrease out of constrained entropy maximisation can be taken as a feature of self-organisation, in that no loop exists if the system joins its environment. Interesting results are given on the role of joint entropies in control in \cite{touchette}, which appear consistent with Theorem \ref{joint-entropy-thm}.





Finally, on the subject of this coupling, note that blanket states play an auxiliary role in the statement and ultimate proof of Theorem \ref{max-ent-FEP-thm}. The Markov blanket itself is subsumed under the constraints determining what states are system-like and what states are not. This has other useful consequences, like the idea of time-varying constraints generalising the FEP to systems with wandering Markov blankets, which was discussed in Section \ref{constraints-as-blankets}. Whilst equivalent, the different viewpoint sheds some light on the nature of blankets\textemdash the same idea, that definitional constraints on system dynamics effectively subsume a Markov blanket, appears in a slightly different fashion in \cite{rorot}. 

\subsection{On the idea of `as if' inference}\label{as-if-inference-subsec} 

As has been pointed out, throughout the literature (see \cite{Andrews2020} for an account of this idea) as well as this paper, the free energy principle holds in most generality when systems look `as if' they are performing inference against their own system-ness. This paper has implicitly put forth the idea that this is the best summarisation of the FEP: an account of the sufficient conditions for a non-equilibrium steady state to exist for a system. This is where its generality in the theory of random dynamical systems comes from, and its description of complex self-organising systems is ultimately a consequence of more mathematical aspirations. This leads to a challenge, however: the FEP is not necessarily married to the supposed physical content of the FEP. A similar sentiment was raised in Section \ref{constraints-as-blankets}\textemdash we model the existence of a system as the performance of inference on system-ness, ultimately an abuse of model interpretation, enabled by the tautology that a particular system carries evidence of being system-like by existing as the system. 
This has been called `organism-centred fictionalism' in \cite{semantics}, emphasising that it is an account\textemdash or indeed, a model\textemdash of how agents behave, given that the system can be understood as being coupled bidirectionally with an environment. This is in contrast to the framework being strictly an explanation or a mechanism. 

We also revisit important comments in Section \ref{duals-subsection}, where we refer to the use of maximum entropy as a modelling tool, capable of understanding an agent from the outside looking in. This is where a key difference between the viewpoints of maximising constrained entropy and minimising free energy emerges. It does not matter whether the agent is truly performing inference: we are performing inference about how the agent behaves, by placing a probability density over its likely states under a semi-arbitrary but informed definition of system-ness. Suppose for some philosophical reason we wish to say that systems do not intentionally maximise their entropy but are modelled by the principle of maximum entropy. This is in line with previous discussions about systems which do not self-evidence, but simply evidence, e.g., Example \ref{inertness-example}. Maximum entropy obviously accommodates this; this was, in fact, the original point of Jaynes' deployment of maximum entropy in statistical physics. We ought to then ask why it is that the formalism here extrapolates to systems with a greater adaptive sense, such as in Example \ref{stat-adapt-ex} (or even Example \ref{non-stat-adapt-ex} when iterated appropriately). The fact that it is still a model is key. Neglecting to explicitly construct actions\textemdash i.e., those originating from active states of the system\textemdash is consistent with the purpose of maximising entropy as a description of some system. This is not a reinforcement learning model of how agents choose actions, and from the perspective of the heat bath the agent is situated in, there is no policy. That is, the environment is agnostic as to the motivational structure for an agent\textemdash there is simply a normative account of what the system is doing at any given time. This is the adjoint of free energy minimisation. We do away with the generative model that the agent's recognition density matches, place a density $q(\mu)$ on what internal states do\textemdash without reference to why, necessarily, they are doing it\textemdash and this matches some prescribed internal states via the maximisation of entropy. We emphasise once more that the lack of action is not simply an artefact of this description\textemdash it is fully a consequence of it. 

\section{Maximum entropy as the description of a field theory}\label{max-ent-field-section}

Recall our motivation is to develop the FEP as the principle underlying a field theory. We assume that the probability density on $X$ has a sufficiently constructive description under field-theoretic axioms to license an interpretation as a real-valued scalar pre-quantum field theory admitting a coupling to an abelian gauge field. That is, $p(x)$ describes a field with probabilistic degrees of freedom, obeys wave mechanics, and has a gauge symmetry. Intuitively, this is fairly clear: the probability measure on statistical microstates, itself dictated by a classical equation of motion, is indeed a central part of statistical field theory (SFT). We take this in a slightly unconventional direction, and invoke this probability density as a field itself. To that end, we will check some key properties, but leave any fuller axiomatic field theory of probabilistic inference to future work. Since we use classical devices to understand the Fokker-Planck equation, any formal equivalence would be mostly for book-keeping purposes. 






Mathematically, an object-valued field theory is simply an assignment of an object state to every point on some base space. For instance, at its mathematically purest,\footnote{In particular, this is the so-called `algebraic approach' to QFT, and includes the cobordism hypothesis that certain quantum field theories are classified by functors (maps) from cobordisms (space-time evolutions) to vector spaces (containing quantum states) \cite{baez-dolan}.} 
a quantum field theory (QFT) is a function assigning a space of quantum states or operators to points or temporal slices of a space-time manifold \cite{haag-kastler, atiyah}, and the same is true of classical fields \cite{brunetti}. Generically, a field is a section of a vector bundle, mapping each point on an input manifold to a space of possible outputs at that point \cite{nakahara}. 

It is clear that a probability distribution over some state space is one such function. Moreover, its evolution is given by a Lagrangian corresponding to a particular wave equation, and under assumptions of a Wiener process, that wave equation can be understood via a path integral representation. This allows us to treat $p(x)$ as a real-valued, semi-classical scalar field theory over probabilistic degrees of freedom (rather than, say, a measure on $X$).





\begin{remark}\label{reification} The notion of \emph{model reification} as a logical fallacy, introduced in \cite{Andrews2020} and \cite{Andrews2022}, arises from transferring the conceptual content of a model to new domains alongside the model itself. One must in general hesitate to treat equivalent mathematical structures as identical objects extending one or another physical interpretation. Conversely, this allows us to use the fact that common mathematical models do not imply common conceptual content, to forget anything but the relevant mathematical structure in a shared model. Whilst perhaps initially unnatural, this is used to effect throughout physics: consider the virtual particles of Feynman diagrammes or the phonons of condensed matter, or the conceptual differences between thermodynamic and information entropy. Here, the claim is not precisely that anything which does inference is a field theory, but rather, that it \emph{looks like} one and thus can be modelled the same way. Conversely, as it is not a physical field theory, we do not necessarily expect to see fully consistent physical features of this theory. This equivalence of models is hence an identification of structures and not necessarily of interpretations. The physical interpretation of this theory is any case complicated by the threefold role of the constraints: $J(x)$ is simultaneously an observable on $X$ whose expectation is $C$, a constraining potential on the acceptability of states in $X$, and a free choice of gauge over $X$. On the other hand, these roles are context-dependent, and can be understood classically as any function of state being both an observable and a potential, and the addition of arbitrary functions on the boundary of $X$ factorising away in the variation of a Lagrangian. 
\end{remark} 

The behaviour of a field theory generally follows from some optimisation principle: laws like the least action principle map onto the fundamental tendency of systems to take trajectories of minimal cumulative energy usage. We preserve this data when we identify $p(x)$ with a field theory, since probability densities minimise the functional 
\[
\int \ln\{p(x)\} p(x) \dd{x},
\]
i.e., they obey the principle of maximum entropy \cite{jaynes}.

\section{Constraints as gauge symmetries}\label{gauge-symmetry-section}

The geometric portion of these results are an application of more general results in previous work by the author \cite{constraint-geometry}. 
We provide a brief account of those results to motivate the results shown here, but refer the technically-minded reader to the cited paper for proofs of these claims. We also briefly motivate gauge theory itself. The gauge-theoretic results induce a flow on the state space describing the sampling dynamics under a given steady state density. This will allow us to show that, for a specific class of non-equilibrium steady state admitting a potential-based description, we can use the entropy-based mathematics we have established for the inert or equilibrium case.

In previous results, we have taken the idea of an ontological potential seriously, as a geometric feature of a system. A potential function is something that appears to shape the flows of a dynamical system in a state space, as if that space were curved. The results indicated begin from the viewpoint that there ought to be a duality between deformations of the geometry of a state space and constraints on the dynamics in that state space, in such a way that maximum entropy translates between the two. Since gauge theory encodes the way that dynamics in a state space reflect the geometry of the state space, the equivalence seems natural; indeed, it is possible to give a new interpretation to the principle of maximum entropy using features of classical gauge field theories. 

\subsection{A review of gauge theory}

In keeping with the attitude in the introduction, the aim is to situate maximum entropy\textemdash and hence the FEP\textemdash as the principle dictating how a geometric foundation determines a dynamical theory. Gauge theory is an especially prescient example. It has a rich geometric framework already in place, in terms of iterated inference on statistical manifolds, where connections to gauge theory \cite{amari, sengupta, sengupta2} have already been made. Likewise, separate connections to gauge forces have appeared in the FEP literature \cite{maxwell2018}, and the Helmholtz decomposition used in the FEP have their roots in the geometric features of simple gauge theories \cite{graham, constraint-geometry}. Going beyond these initial examples, most everything\textemdash from things as small as strings, to things as large as black holes\textemdash and in particular, the statistical mechanics of condensed matter theories \cite{lee}\textemdash have gauge symmetries or connections to gauge theory. That inference or rudimentary forms of cognition should be free of the explanatory scope of gauge theory is unlikely, and certainly unsatisfying. Moreover, many gauge symmetries arise in a certain sense out of constraints on the features of a theory \cite{lee, henneaux, levin-wen, gitman, witten}. It seems appropriate that we should be able to derive some insights for the FEP from gauge theory.

A matter field is a physical field described by an action, as in Section \ref{max-ent-field-section}. A gauge theory is characterised by a matter field theory whose action, or the variation of that action, is invariant under a certain quantity: this quantity is called a gauge \cite{rubakov}. A gauge is a function on space-time coupled to the matter field, in the same way as a potential function couples to the dynamics of a system. A choice of gauge induces a gauge field, akin to the gradient of the gauge potential, keeping track of how that choice of gauge varies across space-time. The geometric formulation of a classical gauge field is via a generalised function called a section, corresponding to a choice of gauge and a matter field configuration. The corresponding generalisation of the derivative of a function is encapsulated in a connection, which gives us a notion of tangent vectors for the image of a section \cite{nakahara}. It follows that gauge fields are connections in a principal bundle, spaces which give rise to such generalised functions.

We take the principle of constrained maximum entropy as a gauge theory, under $p(x)$ a matter field and $J(x)$ a gauge field. This symmetry is rooted in Jaynes’ original claims. Prosaically, the fact that there exists a large class of systems which are all described by maximum entropy, with constraints that can be arbitrarily fixed to produce the particularities of specific systems, makes these constraints a gauge symmetry. We will briefly describe this in more detail, but for full technical constructions, we again refer the reader to \cite{constraint-geometry}. 

Gauge invariance of $S[p]$ is essential to the construction. This is easily shown in analogy to gauge theories like quantum electrodynamics: a short calculation shows that a gauge transformation, a particular sort of change in $J(x)$ compatible with both an assumed coupling of $J(x)$ to $p(x)$ and the geometric structure of the problem, factorises away at some point in the calculation of $p(x)$. This yields an arbitrary choice of gauge as well as an arbitrary change in that choice, leaving the action invariant under any gauge. In contrast, the field equation (here, the probability density) is typically gauge covariant, being expressed in a particular choice of gauge. The interaction with $p(x)$ as $J(x)$ varies is given by parallel transport with respect to $J(x)$, which generalises a dynamical system by means of a covariant derivative. The specific technology of gauge theory, and in particular, relating $J(x)$ to a gauge field coupled to $p(x)$, is motivated by our desire for something that allows us to discuss how varying one quantity varies another in a well-defined way. This is given by the covariant derivative, whose solution is termed parallel transport.

The content of Theorem 1 in \cite{constraint-geometry} is that the solution to this dynamical system, parallel transport, is the stationary point of the entropy functional under a given choice of gauge. This is readily seen by the equation for parallel transport of a field $p(x)$ in a gauge field $\partial_i J(x) \dd{x^i}$, which is 
\[
\partial_i \, p(x) = -\partial_i J(x) p(x)
\]
when expressed in $\R^n,$ $i \in \{1, \ldots, n\}$. Like a separable ODE, this integrates to
\[
-\ln\{p(x) \} - J(x) = 0,
\]
which is the stationary point of entropy. Whilst initially trivial, it reveals a subtle gauge-theoretic structure to our problem, by the nature of parallel transport.

\begin{defn}
A gauge theory consists of a space of matter field states $E$, a space of choices of gauge $P$, and a base space $X$ (generally a space-time manifold like $\R^{3,1}$). The spaces $E$ and $P$ are fibre bundles, copies of the space of possible states at every point in the base space. The bundle $P$ is associated to $E$ such that a choice of points in $P$ frames the expression of points in $E$, and changes in that choice re-frame points in $E$. A choice of points in a bundle is called a section, a sort of internal function assigning bundle points to base points; hence a section defines a field theory. The fact that the matter field is frame-dependent is where gauge covariance comes from. The space that $P$ collates is a symmetry group that acts invariantly on the action of the theory, reproducing invariance. In a bundle there exists a connection that tells us how a section varies across the base, leading to a conceptualisation of the derivative in a bundle. The gauge field is the pullback of the connection to the base, reproducing the effects of the choice of gauge on the evolution of fermionic particles on space-time.
\end{defn}

As a corollary to the above definition, we can construct a gauge theory for maximum entropy: we take the choice of gauge as $g(x) = \exp{J(x)}$, which acts on an exponential density $p(x)$ by multiplication to re-frame it. The local connection $\omega^g_i$ is expressed as $g^{-1}(x) \partial_i g(x)$ in the basis $e^i \partial_i$, such that $\omega^g_i = \partial_i J(x)$. This defines our gauge field. The effect of the gauge field on the evolution of points $p(x)$\textemdash where we mean the application of $p$ to a point $x$, such that this is a scalar changing over space-time\textemdash is given by parallel transport. 

Parallel transport is a special case of motion in a gauge theory which follows the path prescribed by the gauge field exactly, in the sense that when a particle moves in parallel fashion, there is no force acting on particles deflecting them from parallel paths. We can also have motion that experiences a gauge force by curving away from these parallel paths. This splitting of dynamics into trajectories which are parallel to the base\textemdash termed horizontal paths\textemdash and trajectories which curve upwards or downwards, like charged particles\textemdash vertical paths\textemdash is characteristic of a gauge theory \cite{bleek}. In the FEP, it takes on a renewed meaning, which we shall see below. Using a gauge-theoretic and maximally-entropic formulation of the splitting, we can explicitly construct the attractor over states which we sought in earlier results, and, comment on the Helmholtz decomposition in itinerant or life-like systems. 



\subsection{Sample paths in stationary stochastic processes}\label{sample-paths-sec}

Thus far we have focussed on a relatively simple theory: that of an effectively equilibrium system. In particular, we have not covered the dynamical sample paths admitted by the evolution of states under $p(x)$, which are not, themselves, stationary; in fact, invoking maximum entropy as a model of a system in Section \ref{as-if-inference-subsec}, we have intentionally \emph{ignored} sampling dynamics. We can no longer avoid this, given the content of parallel transportation, but we still constrain ourselves to this simple case. 

We begin by revisiting results from \cite{constraint-geometry} in the context of stochastic systems which self-organise. Using the results on estimation as a gauge force, inference can be understood as weighting the states of a process in such a way that inference converges to a stationary distribution over some data by constraining inference to the set of most likely states. We begin the argument by giving a formal statement of the idea that a system has a system-like attractor. We are then able to prove that the gauge theory induced by the constraints enforces the convergence of sample paths to that attractor under strong constraints, and that this builds a maximum entropy probability distribution over states by purely geometric considerations. In the Gaussian case, an attractive intuitive picture is that of a funnel, guiding paths to a low-constraint region, such that this region is an attractor where states are observed with high probability. 
Under this interpretation, we return to the idea of 
the constraints being a potential function, guiding the dynamics of a process. In fact, we assume $p(x)$ is a Gaussian serving as the Laplace approximation to an underlying density. We also discuss splitting behaviours of stochastic dynamics under this gauge potential, referring implicitly to Proposition 4 of \cite{constraint-geometry}.

The general, formal definition of a `geometry,' a motif dating back to Klein's \emph{Erlangen} programme and which features heavily in work as recent as Lurie's derived algebraic geometry \cite{lurieDAGV} and Freed and Hopkins' $H_n$-structures \cite{freed-hopkins}, is that of a topological space equipped with some additional structure, especially that which derives from a group or a group action. A Euclidean geometry, for instance, is an $n$-dimensional real space equipped with a Euclidean metric and isometries thereof. A Riemannian geometry is a smooth manifold equipped with a Riemann structure. Previously, we have introduced the idea of a \emph{constraint geometry} as a geometric formulation of the covariance of probabilities and constraints in maximum entropy. The fixing of constraints in maximum entropy induces a potential function that affects flows in that space, reflecting an extra structure in what states are accessible by a flow and what states are not. We offer the following semi-formal definition to that end:

\begin{defn}\label{CG-def}
A constraint geometry is a topological setting in which dynamics in some space interact with a constraint function in that space determining the shape of any such flow. Constraint geometries have a natural interpretation as the result of a gauge fixing procedure in a manifestly $G$-covariant field theory for some Lie group $G$, and hence admit a formulation in terms of gauge theory. As such, a constraint geometry is a principal bundle structure over a space of interest with a given choice of connection, and an associated bundle giving some ensemble property of the possible flows.
\end{defn}

Notice what we have done without much further elaboration: claim that constraints tell us what states can and cannot be accessed, rather than what states can be accessed and with what probability. Intuitively, we can relate the idea of sampling under a density to the frequencies of sampled states, as though histogramming the occupation time of our system. Under very strong constraints this means that flows will occupy constrained states very briefly, if at all, which returns us to the idea of the locally ergodic measure constructed in Theorem \ref{ergodic-thm}. In fact, we can formalise this using aspects of ergodic dynamical systems theory. 

\begin{defn}\label{time-ergodic-defn}
An ergodic probability measure implies that the proportion of time spent in an area of state space $A$ is equivalent to $P(A)$, i.e., 
\begin{equation}\label{time-ergodic-eq}
\lim_{T\to\infty} \frac{1}{T} \int_0^T \mathbf{1}_A(x_t) \dd{t} = P(A)
\end{equation}
for $\mathbf{1}_A$ an indicator function on states $x \in A$ and $x_t$ a state at time $t$.
\end{defn}

Definition \ref{time-ergodic-defn} leads to the celebrated equivalence of temporal and spatial averages for ergodic measures, due originally to Birkhoff. Here, it allows us to claim that an adequately system-like system which visits all system-like states defined under some set of constraints will almost surely be observed in that region $A$. Correspondingly, in the limit $T\to\infty$ when system-like events become typical, we can interpret the locally ergodic measure for a constrained system as a demand that the system spend negligible time outside of $A$.

\begin{theorem}\label{funnel-thm}
Let the open set $A$ be some region of the state space. The following statements are true, and under Theorem \ref{ergodic-thm}, arise from a constraint geometry:
\vskip0.5em
\begin{thmnum}
    \setlength\itemsep{0.5em}
    \item Approximate Bayesian inference arises from a gauge force acting on a locally ergodic system in the direction of an optimal control parameter.\label{statement-1}
    \item A locally ergodic system occupies any state in $A$ with probability $\exp{-J(A)}$, which is equivalent to exploring $A$ in parallel fashion.\label{statement-2}
    \item For a locally ergodic system, there exists an attractor $A$.\label{statement-3}
\end{thmnum}
\vskip0.5em
\end{theorem}
\begin{proof}
To prove Statement \ref{statement-1}, we need only note that, on a Gaussian probability density, under a gauge potential, there exists a vertical flow towards the expected internal state. By Theorem \ref{constraint-is-potential-thm}, this potential is a constraint on the maximum entropy probability density, which is approximate Bayesian inference by Theorem \ref{max-ent-FEP-thm}. As such, maximising $p(x)$ to find the expected internal state in $A$ is a vertical flow, which we identify as the flow of charged particles under a gauge force. To prove Statement \ref{statement-2}, we recall results from \cite{constraint-geometry}; namely, that under a gauge potential, there exists a horizontal flow of equiprobable states which explores that locus of equiprobable states. The set of all such level sets is $\exp{-J(A)}$. Now we prove Statement \ref{statement-3}. Firstly, for a locally ergodic system whose occupation of system-like states has measure zero, the solenoidal flow degenerates on the complement of $A$ such that states are confined to $A$. By Poincar\'e recurrence, the set of points in $A$ that escapes $A$ has measure zero if $P(A) > 0$, such that states return to this confining region; since $A$ is an open set, this is finally a trapping region for the system, and thus an attractor. Now we can prove that all three statements are implied in a constraint geometry. Statements \ref{statement-1} and \ref{statement-2} are clearly equivalent to the existence of a gauge potential, without which the splitting of flows under constraints on inference does not exist. Under Theorem \ref{ergodic-thm}, our use of Poincar\'e recurrence necessitates a constraint geometry, and is equivalent to strong constraints on $\F \setminus A$. As such, Statement \ref{statement-3} holds if and only if we maximise entropy in a constraint geometry. This proves the primary claim.
\end{proof}

We take a moment to comment on the results in Theorem \ref{funnel-thm} more fully\textemdash the region of high probability $A$ constructed in this result is the system-like attractor that we initially sought, in Section \ref{bayes-mech-prelim-sect}. The intuitive idea is that the FEP says our model of a system should be that the system's evolution returns to this region of system-like states, in virtue of it being such a system. The fact that the system samples from these states is described by the steady state density over $A$, which here has become an ergodic density for a trapping region $A$. When constraints $J(x)$ exist such that the complement of $A$, $\F \setminus A$, has measure zero, we can rigorously say that the system spends no time in that region\textemdash otherwise, we must enforce a semi-arbitrary variance cutoff defining what counts as an instance of that system and what is leaving system-like states.


\begin{example}\label{stone-attractor-example}
Consider a dissipative dynamical system and a given blanket state for it, such as a stone for a given weather condition. A faithful model of a stone must constrain itself such that it proceeds directly to the expected internal state given that blanket state; less normatively, we can expect a stone to be defined as an inert system, which definitionally, does not have the ability to vary its states. As such, the ontological potential of a stone (conditioned on a blanket state) is highly constrained about that expected internal state. In this case, the solenoidal component of the flow degenerates, whilst the gauge force we have referred to enforces the thing-ness of the stone by driving it to exist at $\hat\mu_b$. Here the attractor given a blanket state \emph{is} $\hat\mu_b$, and the conditional system-like density is the Dirac measure on $\hat\mu_b$; the full set of system-like $\mu$, $A\subset X$, is the set of all such possible stone-like states, with individual probabilities based on the constraints we place on what it means to be a stone. The systemic constraints on a stone would likely be constant within $A$, since there is no preference to be expressed, and hence would produce uniform probabilities over $A$ and zero probabilities elsewhere. The resulting probability density could be updated were we to observe some states more often than others, which would place further accuracy constraints on our model in addition to constraints definitional of the system.
\end{example}

In particular, such constraints could include a hard cut-off of temperature at whatever the stone's melting point is, as well as a constraint on the internal energy of the molecular arrangement of the stone, enforcing its structure as, say, a silicate mineral. Some of these states will be observed less often, suggesting we should change the shape of the constraints until we get the desired sampling relation on $A$ yielded by \eqref{time-ergodic-eq}. This could be predicated on certain expected internal states being observed more often because their blanket states are observed more often, changing the density $p(b)$ that we marginalised over in Theorem \ref{max-ent-FEP-thm}. 



A consequence of formulating these results as parallel transport is that the way this sampling is enforced is by splitting the dynamics under such a descriptive steady state density:

\begin{corollary}\label{grad-cor}
Under a choice of gauge $g(x) = \exp{-J(x)}$ and an induced field equation $p(x) = \exp{-J(x)}$, a solenoidal flow over the state space exists. 
\end{corollary}
\begin{proof}
For a gauge theory where $g(x) = p(x)$ and a differential acting in the standard basis of $\R^n$, the parallel transport equation satisfies the following logarithmic derivative:
\begin{equation}\label{par-trans-surprisal-grad}
\partial_x p(x) = -\partial_x \ln\{ p(x) \},
\end{equation}
conventionally written as $g^{-1} \dd{g}$. Moreover, Statement \ref{statement-1} holds in the vertical direction where this gradient is non-zero, and Statement \ref{statement-2} holds in the horizontal direction where this gradient is zero. Since the lift of a path $\varphi_t = \{x_1, \ldots, x_n\}$ is given by $p \circ \varphi$, the pullback to $X$ under quadratic constraints gives circular paths for horizontal $\varphi$ and flows towards $\hat x$ for vertical $\varphi$. This yields flows in $X$ along the surprisal.
\end{proof}

As an addendum to the above corollary, note that the derivative on the left-hand side, $\partial_x p(x)$, measures how the evaluation of the section $p$ at a point changes; hence it is analogous to a particle moving on the probability density treated as a hypersurface. As such, under Theorem \ref{funnel-thm}, Corollary \ref{grad-cor} proves that particles which perform approximate Bayesian inference perform a gradient descent on surprisal, as suggested in \cite{afepfapp, parr, dacosta}. See also Example 1 of \cite{constraint-geometry}. 

Note what it does not prove: that those particles seek the mode $\hat x$. They merely match it. Moreover, note that the velocity of the flow\textemdash the parameterisation of the isocontours, in other words\textemdash is arbitrary. Supposing a matrix operator of the form given in \cite{afepfapp, dacosta, dacosta2} acts on the flow to yield a well-defined velocity along a curve recovers those results. However, this must be specified beforehand. Reintroducing a Lagrange multiplier to \eqref{par-trans-surprisal-grad} and choosing the value of $\lambda$ based on some physical data to be matched may be a more principled way of generating those matrices, but a worked example should be provided in the future. 

As was mentioned, the paths we describe are relatively simple, given that (by construction) they converge to a stationary attractor. A proper theory of paths, which extends to moving attractors and even focusses on the paths themselves\textemdash and thus, describes adaptive, truly complex systems\textemdash requires a broad generalisation of these results, but one which is within reach.

We now discuss an application of these results to non-equilibrium situations. Some connections to the splitting introduced above are mentioned.

\subsection{Regarding non-equilibria}\label{non-eq-subsection} In this section, we are able to gesture towards an application of these results to non-equilibrium regimes. Thus far we have been able to show that the relationship between the technology of the FEP and the nature of equilibrium dynamics rests on fairly solid ground.

We have previously referred to the lack of `effective' energy flows in non-equilibrium steady state. It is apparent that, due to stationarity playing a fundamental role in both equilibrium and non-equilibrium steady state, the only difference between these systems is the presence of some underlying flux. 

\begin{example}
Following Example \ref{stone-attractor-example}, when the locally ergodic measure for a system which performs approximate Bayesian inference is $\delta(x - \bar{x})$, and the resultant attractor is a fixed point $\bar{x}$, the horizontal flow degenerates and there is no solenoidal component. Thus, we often need to make a choice as a modeller as to whether or not it makes sense to describe a system with some spread in its state space. For this reason, an underlying horizontal flow is more like a non-equilibrium condition than an equilibrium one; in the limit, equilibrium systems will go to $\bar{x}$ under any sort of dissipation. Indeed, if the horizontal flow has a meaningful sense of direction\textemdash if a flow in one direction is preferable to a flow in another, and as such, if forward trajectories are more probable than their inverses\textemdash then detailed balance is broken, making this a non-equilibrium steady state. However, as we mentioned, this is strictly extra (physical) data about the horizontal flows. Correspondingly, maximum entropy can accommodate both under appropriate constraints.
\end{example}

Coarse-graining non-equilibrium systems at steady state, where those fluxes cancel out, implies that non-equilibrium steady state \emph{looks like} equilibrium steady state with spread and can be modelled the same way. Ignoring certain physical degrees of freedom, we can use maximum entropy to treat non-equilibrium steady state and still derive insights. This is fully consistent with maximum entropy as a model (cf. Sections \ref{duals-subsection} and \ref{as-if-inference-subsec}). Our model of the system's characteristic probability density is agnostic to the sampling dynamics and actions taken by the system under that density: we simply do not evaluate \emph{how} a non-equilibrium steady state is maintained, only \emph{whether} it is. As such, the underlying sampling dynamics and physical degrees of freedom can be ignored to get an effective equilibrium. 
This is another sleight of hand that is necessary to extend our simple picture to the less trivial heart of the FEP, but which is possible in any case. 

\begin{remark}[limitations of effective equilibria]\label{limitations-rem}
It should be noted that, realistically, this framework is only insightful near equilibrium, where the non-zero probability flux due to non-equilibrium can actually be neglected \cite{zia, england3}. Consider also that we do not want divergences in entropy, but only presume it is maximised within what is allowable by constraints. As such, the work done by underlying flows cannot be arbitrary\textemdash it must be bounded\textemdash another signature of near-equilibrium. 
On the other hand, assuming the system has reset its energy levels to whatever is induced by the constraint is effectively designing a new equilibrium for the system \cite{udo}, which is what we do as modellers who coarse grain our observation of the system's maintenance of a stationary state. Some further technical conditions should be stated: by assuming our system's sampling dynamics are related to entropy production and reflect a coupling to an environmental heat bath, this requires at least local detailed balance, purely for our model to be valid \cite{maes}. 
Luckily, we lose very little by these simplifying assumptions, in that the production of entropy as a characteristic of life (cf. Remark \ref{2nd-law-rem}) is a hallmark of the non-equilibrium steady states of even complex adaptive systems \cite{rovelli, kai}.
\end{remark}


On this basis, we have suggested that this construction applies to systems with out of equilibrium dynamics which can be described by a potential. An example of this, and a justification of the use of maximum entropy on that system, is constructed in this section. In particular, we prove an effective equilibrium theorem below, which constructs a generic system like Example \ref{stat-adapt-ex} that is described by maximum entropy, thereby proving such a class of system exists.


\begin{prop}\label{stoch-entropy-prop}
Non-equilibrium systems with a steady state as an initial condition maximise entropy at any other steady state.
\end{prop}
\begin{proof}
Define a stochastic entropy functional as $s[x] = -\ln\{ p(x(t)); \kappa(t))\}$, which is simply the surprisal on a trajectory for some control parameter $\kappa(t)$. 
At steady state, the stochastic entropy can be averaged into the Shannon entropy, since any $p(x(t)) = p(x)$. Now suppose $p(x; \kappa) = \exp{-s(x)}$. This is the stationary point of the Shannon entropy.
\end{proof}

The stochastic entropy in Proposition \ref{stoch-entropy-prop} can be taken, rather straightforwardly, as a constraint that the system ought to meet a particular non-equilibrium steady state and remain there. In the following, we use a different constraint, assuming the potential captures both the driving force towards an equilibrium density and the minimum energy necessary to maintain a density favouring states with statistics out of equilibrium. In so doing, we extend the above proof to systems with control parameters, an important viewpoint on non-equilibrium systems for our purposes. 

\begin{theorem}\label{eff-eq-thm}
Non-equilibrium systems at steady state under a potential function are systems which maximise entropy under a constraint and minimise variational free energy.
\end{theorem}
\begin{proof}
We prove this by constructing an example of such a system.
Take a controlled system in the presence of non-zero flux of energy. 
Suppose the system constrains itself to fluctuate around a mean state $\hat x$. Moreover, suppose $\hat x \neq \hat x_{\text{eq}}$. This system is at non-equilibrium. Finally, suppose that $\hat x$ is both a fixed point and a sufficient statistic for the sampling density, such that it is stationary. There is a potential function $E(x-\hat x)$ measuring the deviation from $x^*$ such that on average, $E(x-\hat x) = 0$. This potential function is a coarse-graining of the underlying energy flux induced by drives and dissipation in the system. These fluxes increase entropy production; as such the system is a non-equilibrium steady state which maximises entropy under the constraint $\E[E(x-\hat x)]=0$. By constraining itself to $\hat x$, this system is equivalent to a system which engages in approximate Bayesian inference (Theorem \ref{max-ent-FEP-thm}). 
\end{proof}

At the very least, $E$ can be considered a quadratic form, and thus the Laplace approximation of a non-equilibrium system representing the distance from the mode. Consistent with the formalism laid out in \cite{udo}, a non-equilibrium steady state $p(x;\hat x)$ assumed to have relaxed from an initial non-equilibrium steady state\textemdash rather than a transient state\textemdash admits a potential based on a control parameter $\kappa$, which is in general equivalent to $-\ln\{p(x; \kappa) \}$. Invoking Theorem \ref{max-ent-FEP-thm} allows us to pose this as a problem of maximising constrained entropy, and dually, one of minimising free energy. In \cite{niven2, niven1} a similar potential-based description of a non-equilibrium steady state, formulated in terms of maximising entropy without explicit reference to the underlying fluxes, is given. 





We raise this problem now to provide some comments about the solenoidal flows written of in the FEP. The results from the previous section combined with Theorem \ref{eff-eq-thm} constitute a proof that, for a certain class of non-equilibrium systems, a solenoidal flow exists. The generality of this decomposition for stationary systems has been proven rigorously before \cite{betancourt}, and it is formulated in detail in \cite{dacosta, dacosta2}.\footnote{It has been known for some time that such a decomposition exists for Stratonovich stochastic differential equations, perhaps first being described in \cite{graham}; the papers and results discussed here focus on its applications to the FEP. The author thanks Lancelot Da Costa for suggesting this reference.}

Using the splitting of dynamics induced by a gauge-theoretic structure on our state space, we are able to state that a solenoidal flow on states exists. The decomposition of flows on the surface $(x, p(x))$ induces an equivalent decomposition of the flows on the $X$, the state space, by the pullback one-form (the local gauge field). It is in this manner that we reproduce solenoidal and dissipative dynamics, which explore and drive to a fixed point, respectively; this accommodates the flow decomposition referred to in \cite{afepfapp}. The spread of the density is seen as a fluctuation-based, exploratory flow, whilst the dissipative flow is a mode-matching flow which corresponds to evidencing, or the control that maintains a non-equilibrium steady state. In inert cases this fixed point is nothing quite special, and simply represents the average state of the system contingent on the average external state\textemdash such a mode exists by evidencing, but has no special content, since the system does not self-evidence. Contrast this with other systems that pursue a mode that corresponds to a desired belief, and we recover a sort of adaptivity.



Note what we have not claimed: that there are any dynamics to the expected states. Moreover, by ignoring the form of flows on the state space, we are not able to reproduce key results of the monograph \cite{afepfapp}, which are non-trivial \cite{aguilera2021} assumptions about the decomposition of flows under the FEP. What results we have been able to reproduce hold identically at equilibrium and non-equilibrium in virtue of their connection to stationarity, and have little connection to the underlying flows or couplings; they simply arise from a natural gauge-theoretic interpretation of the FEP. Parameterising the change in the density on a level set gives us curves of arbitrary velocity in \eqref{par-trans-surprisal-grad}, but there is still no explicit connection to the sampling happening under that density. It is interesting that the solenoidal flow surely exists, but approximate Bayesian inference is possible without it, as in Theorem \ref{ABIL-thm}. From that perspective, that the existence and interpretation of this flow is noted in \cite{afepfapp} with the fact that this flow does not necessarily imply FEP-type inference in \cite{aguilera2021} may be resolved. 

There are some further caveats to these results. For instance, the decomposition is also not unique\textemdash the choice of horizontal tangent space can be reparameterised or chosen differently. This is consistent with the literature, in the sense that there is a large equivalence class of dynamics that fall under the same non-equilibrium steady state density and hence the potential for any such density is non-unique. The splitting exists, but the flows underlying that splitting are arbitrary. That is, just knowing the non-equilibrium steady state is not sufficient to produce unique forms for the flows. 
In fact, this too is consistent with modelling philosophies about using the FEP: we can't access precise information about internal states of a system, we can only model what it does at the level of ensemble statistics. What we can say is that the flows go towards that region (Theorem \ref{funnel-thm}) and that this flow, at some ensemble level, looks like it consists of solenoidal and dissipative components\textemdash these local vector fields must go towards the trapping region, and that there is a splitting of those fields. 


It is possible that some of these issues will be addressed in a more dynamical formalism than simply maximising entropy, such as maximising path entropy, which\textemdash by Proposition \ref{stoch-entropy-prop}\textemdash is already an attractive generalisation to truly life-like systems.

\section{Conclusion}

This paper has reformulated ideas fundamental to the free energy principle in the language of maximum entropy and gauge theory. We have established that, at minimum, a slightly simplified version of the FEP is a sound mathematical principle, the implications of which are verifiable, and which have an additional gauge-theoretic interpretation.

Throughout, we have leveraged a few extensions or relaxations of more traditional concepts. Namely, it would that the FEP satisfies properties like ergodicity, but only locally so; maximises entropy, but only under constraints; applies at non-equilibrium, but only because it applies to stationary systems in a scale-free fashion. The FEP rests on several such results\textemdash places or equations where it bends a definition just enough to work in a rigorous framework. This could be taken as a sign that the FEP maps a territory lying towards the edge of what can be formalised in conventional mathematics. Correspondingly, it is apparent that the FEP is\textemdash at some basic level of description, and as suggested here, certainly more generally\textemdash essentially correct. Rather than being riddled with `global errors' in conceptualisation and application, as some critics have suggested, its failings are limited to `local errors' in the precision of mathematical statements and techniques; these reflect a global correctness, but a technical subtlety, to the theory\textemdash and with a bit of work, they are easily smoothened out.

Recalling the rich history of physically-inspired mathematics and mathematically formulated physics, the framework laid out in this paper builds the FEP a concrete, formal foundation, with some results being of more general interest to the mathematical theory of inference and stochastic dynamical systems. In the future, this foundation will allow easier and more effective generalisations of the mathematics contained in the FEP, to describe more complicated systems just as rigorously. 

\bibliographystyle{alpha}
\bibliography{main}

\end{document}